\documentclass[aps,nofootinbib,showpacs,showkeys,preprintnumbers,amsmath,amssymb]{revtex4}

\usepackage{amsmath,amsfonts,amssymb,amsthm,bm,bbm,bbold,braket,color,dsfont,fancybox,hyperref,
srcltx,subfigure,ucs,yfonts,cancel,xfrac,verbatim,xcolor}
        
\usepackage{float}

\usepackage{mathtools}
\usepackage[T1]{fontenc} 
\newcommand{\beq}{\begin{equation}}
\newcommand{\eeq}{\end{equation}}
\newcommand{\bea}{\begin{eqnarray}}
\newcommand{\eea}{\end{eqnarray}}
\newcommand{\beas}{\begin{eqnarray*}}
\newcommand{\eeas}{\end{eqnarray*}}
\newcommand{\bi}{\begin{itemize}}
\newcommand{\ei}{\end{itemize}}

\usepackage{amsmath,amsfonts,amssymb,amsthm,bm,bbm,bbold,braket,color,dsfont,fancybox,hyperref,
srcltx,slashed,ucs,yfonts,cancel,xfrac,verbatim,xcolor}

\usepackage[inline,shortlabels]{enumitem}
\usepackage{nicefrac}
\usepackage{multirow}

\DeclareMathAlphabet{\mathpzc}{OT1}{pzc}{m}{it}
\definecolor{gold}{rgb}{1,0.8,0}
\definecolor{nara}{rgb}{1,0.4,0.1}
\definecolor{goldo}{rgb}{1,0.7,0}
\definecolor{greeno}{rgb}{0,0.8,0}
\def\bes{\begin{subequations}}
\def\ees{\end{subequations}}
\def\be{\begin{equation}}
\def\ee{\end{equation}}
\def\bea{\begin{eqnarray}}
\def\eea{\end{eqnarray}}
\def\ba{\begin{eqnarray}}
\def\ea{\end{eqnarray}}
\def\bear{\begin{array}}
\def\eear{\end{array}}

\newcommand{\bpm}{\begin{pmatrix}}
\newcommand{\epm}{\end{pmatrix}}
\newcommand{\BM}{\left(\begin{array}}		
\newcommand{\BMC}{\left[\begin{array}}		

\newcommand{\EM}{\end{array}\right)}		
\newcommand{\EMC}{\end{array}\right]}		

	% ...

\newcommand{\com}[1]{}
\newcommand{\K}{\mathcal{K}}

\begin{document}

\title{Probing heavy neutrino oscillations in rare W boson decays} 

\author{Gorazd Cveti\v{c}$^{1}$}
\email{gorazd.cvetic@usm.cl}
\author{Arindam Das$^{2}$}
\email{arindam@kias.re.kr}
\author{Jilberto Zamora-Sa\'a$^{3}$}
\email{jilberto.zamora@unab.cl}

\affiliation{$^1$Department of Physics, Universidad T\'ecnica Federico Santa Mar\'ia, Valpara\'iso, Chile.}
\affiliation{$^2$School of Physics, KIAS, Seoul 02455, Korea.}
 \affiliation{$^3$Departamento de Ciencias F\'isicas, Universidad Andres Bello,  Sazi\'e 2212, Piso 7,  Santiago, Chile.}
\begin{abstract}
In this work, we study the lepton number violating W boson and top quark decays via intermediate on-shell Majorana neutrinos $N_j$ into three charged leptons and a light neutrino. We discuss the neutrino oscillation effects present in the decay due to the small mass gap between the heavy neutrino states ($\Delta M_N \sim \Gamma_N$). We focus on a scenario that contains at least two heavy Majorana neutrinos in the mass range $\sim 2-80$ GeV. The results indicate that the modulation of the branching ratios as a function of the distance between the vertices may be detected in a future experiment such as High-Luminosity Large Hadron Collider. As a secondary result, the CP-violating phases could be explored.
\end{abstract}

\keywords{Heavy Neutrino Oscillations, Lepton Number Violation, LHC.}

\maketitle

\section{Introduction}
\label{s1}

The neutrino oscillation \cite{Pontecorvo:1967fh}, Baryonic Asymmetry of the Universe (BAU) \cite{Canetti:2012zc} and Dark Matter (DM) \cite{Murayama:2007ek} are clear indications of physics beyond the standard model (SM). The discovery of the neutrino oscillation \cite{Neut1,Neut2,Neut3, Neut4, Neut5, Neut6} has proved the existence of the tiny neutrino mass and the flavor mixing which is one of the missing pieces of the SM. The seesaw mechanism is the natural way which leads to the dimension five Weinberg operator \cite{Weinberg:1979sa} to naturally understand the existence of the small neutrino mass. As a result we can extend the SM using SM-gauge singlet Right Handed Neutrinos (RHNs) to naturally explain the tiny neutrino mass using the seesaw mechanism \cite{seesaw0,seesaw1,seesaw2,seesaw3,seesaw4,seesaw5,seesaw6}. In this model we introduce SM gauge-singlet Majorana RHNs $N_R^{\beta}$ where $\beta$ is the flavor index. $N_R^{\beta}$ couples with the SM lepton doublets $\ell_{L}^{\alpha}$ and the SM Higgs doublet $H$. The relevant part of the Lagrangian is
\bea
\nonumber
\mathcal{L} = -Y_D^{\alpha\beta} \overline{\ell_L^{\alpha}}H N_R^{\beta} 
                   -\frac{1}{2} M_N^{\alpha \beta} \overline{N_R^{\alpha C}} N_R^{\beta}  + H. c. .
\label{typeI}
\eea
After the spontaneous EW symmetry breaking
   by the vacuum expectation value (VEV), 
   $ H =\begin{pmatrix} \frac{v}{\sqrt{2}} \\  0 \end{pmatrix}$, 
    we obtain the Dirac mass matrix as $M_{D}= \frac{Y_D v}{\sqrt{2}}$.
Using the Dirac and Majorana mass matrices 
  we can write the neutrino mass matrix as 
\bea
\nonumber
M_{\nu}=\begin{pmatrix}
0&&M_{D}\\
M_{D}^{T}&&M_{N}
\end{pmatrix}.
\label{typeInu}
\eea
Diagonalizing this matrix we obtain the seesaw formula
 for the light Majorana neutrinos as 
\bea
\nonumber
m_{\nu} \simeq - M_{D} M_{N}^{-1} M_{D}^{T}.
\label{seesawI}
\eea
 
The searches of the Majorana RHNs can be performed using the trilepton final state and decays of rare meson \cite{Das:2012ze,Das:2014jxa,Cvetic:2013eza,Cvetic:2014nla,Cvetic:2015ura,Cvetic:2015naa,Das:2015toa,Das:2016hof,Mejia-Guisao:2017gqp, Milanes:2016rzr,Cvetic:2016fbv, Cvetic:2017vwl}, rare $\tau$ decays \cite{Zamora-Saa:2016ito,Kim:2017pra,Dib:2011hc}, same-sign dilepton plus dijet signal \cite{Chatrchyan:2012fla,Helo:2013esa,Dib:2015oka,Dib:2016wge,Arbelaez:2017zqq}, trileptons \cite{Dib:2017vux,Dib:2017iva}, however, those processes will be suppressed by the squared of the light-heavy mixing parameter $|B_{\ell N}|^2\simeq |M_D M_N^{-1}|^2$ where $M_D$ and $M_N$ are the Dirac and Majorana masses of the seesaw matrix, respectively. 
A comprehensive general study on the parameters of $|B_{\ell N}|^{2}$ have been given in \cite{Das:2017nvm} using the data from the neutrino oscillation experiment \cite{Neut1,Neut2,Neut3,Neut4,Neut5,Neut6}, bounds from the Lepton Flavor Violation (LFV) experiments \cite{Adam:2011ch, Aubert:2009ag, OLeary:2010hau},
Large Electron-Positron (LEP) \cite{Achard:2001qv, delAguila:2008pw, Akhmedov:2013hec} experiments using the non-unitarity effects \cite{Antusch:2006vwa, Abada:2007ux} applying the Casas- Ibarra conjecture \cite{Casas:2001sr}.
At this point we mention that  \cite{Das:2017nvm} has a good agreement with a previous analysis \cite{Rasmussen:2016njh} on the sterile neutrinos. In \cite{Das:2017nvm} we can put bounds on the light- heavy mixing angles using results from different experiments as shown in \cite{ Achard:2001qv,delAguila:2008pw, Akhmedov:2013hec, deBlas:2013gla, BhupalDev:2012zg, Aad:2015xaa, CMS8:2016olu, KamLAND-Zen:2016pfg, Das:2017zjc,Das:2017rsu} considering the degenerate Majorana RHNs.

 In canonical seesaw, $M_N\sim 100$ GeV implies that the Dirac Yukawa coupling $Y_{D}$ between the SM $SU(2)$ doublet leptons, $SU(2)$ doublet Higgs and Majorana heavy neutrino could be $\mathcal{O} (10^{-6})$. Such situation could also be improved if the mass matrices $M_D$ and $M_N$ have specific textures being enforced by some symmetries \cite{Pilaftsis:2003gt, Kersten:2007vk, Xing:2009in,Zamora-Saa:2016qlk} so that a large light-heavy neutrino mixing could be possible even at a low scale satisfying the neutrino oscillation data and allowing to explore the BAU by means of leptogenesis \cite{Chun:2017spz}. Such heavy neutrinos could also be produced at the high energy colliders through the same-sign dilepton plus dijet signal. However, the mass scale of the RHNs is not yet fixed therefore it is considered as a free parameter. Therefore for a heavy neutrino mass in the GeV scale or below the Dirac Yukawa coupling will be small to reproduce the correct SM light neutrino mass.

Through the seesaw mechanism, a flavor eigenstate ($\nu$) of 
  the SM neutrino is expressed in terms of the mass eigenstates 
  of the light ($\nu_i$) and heavy ($N_m$) Majorana neutrinos such as 
\bea 
\nonumber
  \nu \simeq  B_{\ell i} \nu_i  + B_{\ell N} N_m,  
\eea 
where $B_{\ell i}$ is the $3\times 3$ light neutrino mixing matrix being identical to the PMNS matrix at the leading order if we ignore the non-unitarity effects. Whereas $B_{\ell N}$  is the mixing between the SM neutrino and the RHNs, 
  and we have assumed a small mixing, $|B_{\ell N}| \ll 1$.  
Using the mass eigenstates, the relevant terms of the charged current interaction for the heavy neutrino 
  is given by 
\bea 
\mathcal{L}_{CC} = 
 -\frac{g}{\sqrt{2}} W_{\mu}
  \bar{\ell} \gamma^{\mu} P_L   B_{\ell N} N_m  + h.c., 
\label{CC}
\eea
where $\ell$ denotes the three generations of the charged leptons in the vector form, and 
  $P_L =\frac{1}{2} (1- \gamma_5)$ is the projection operator. 
Similarly, the relevant parts of the neutral current interaction is given by 
\bea 
\mathcal{L}_{NC} = 
 -\frac{g}{2 c_w}  Z_{\mu} 
\left[ 
  \overline{N_m} \gamma^{\mu} P_L  |B_{\ell N}|^2 N_m 
+ \left\{ 
  \overline{\nu_m} \gamma^{\mu} P_L B_{\ell N}  N_m 
  + h.c. \right\} 
\right] , 
\label{NC}
\eea
 where $c_w=\cos \theta_w$ with $\theta_w$ being the weak mixing angle. 

In this paper we consider the RHNs having masses within $m_\tau+m_e < M_N < M_W-m_{\mu}$ to study their production from the $W$ decay $W \to N\ell_1^{+}, N \to \ell_2^{+} \ell^{-}_3 \overline{\nu}$ at the LHC where $m_\tau, m_{e}, m_{\mu}, M_W$ and $M_N$ are the tau $(\tau)$, electron $(e)$, muon $(\mu)$, SM $W$ boson and RHN masses respectively. Our paper is arranged in the following way. In Sec.~2 we review the bounds on the RHN mixing angles as a function of $M_N$ relevant with our study. We also include the new existing and prospective limits from the direct searches at high energy colliders. In Sec.~3 we calculate the different decay widths of the RHNs. In Sec.~4 we discuss the possibility of finding such RHNs from the $t$ quark decay. 
 Sec.~5 is dedicated for the conclusions.
\section{Current bounds on the RHN mixing angles}
\label{s2}

The RHN mass $(M_N)$ is a free parameter. Different choices had been considered in the previous literatures such as $M_N < 1$ MeV \cite{deGouvea:2005er} in the eV-seesaw models, however,  the neutrino oscillation data rules out the possibility of the RHNs between $1$ neV $\leq M_N \leq 1$ eV \cite{Cirelli:2004cz, deGouvea:2009fp, Donini:2012tt}. The neutrinoless double beta $(0\nu2\beta)$ decay and the precision measurements of the $\beta-$ decay energy spectra constrain the mixing between the RHN and the electron neutrino between $10$ eV $\leq M_N \leq 1$ MeV. The peak searches of the leptonic decay of the pions and the kaons constrain the mixing between the RHNs and both of the electron and muon neutrinos between $1$ MeV $\leq M_N \leq 1$ GeV. The beam dump 
experiments have searched the RHNs through the decays to put bounds on the mixing angle with all the flavors of the light neutrinos. 

\begin{figure}[H]
\begin{center}
\includegraphics[scale=0.3]{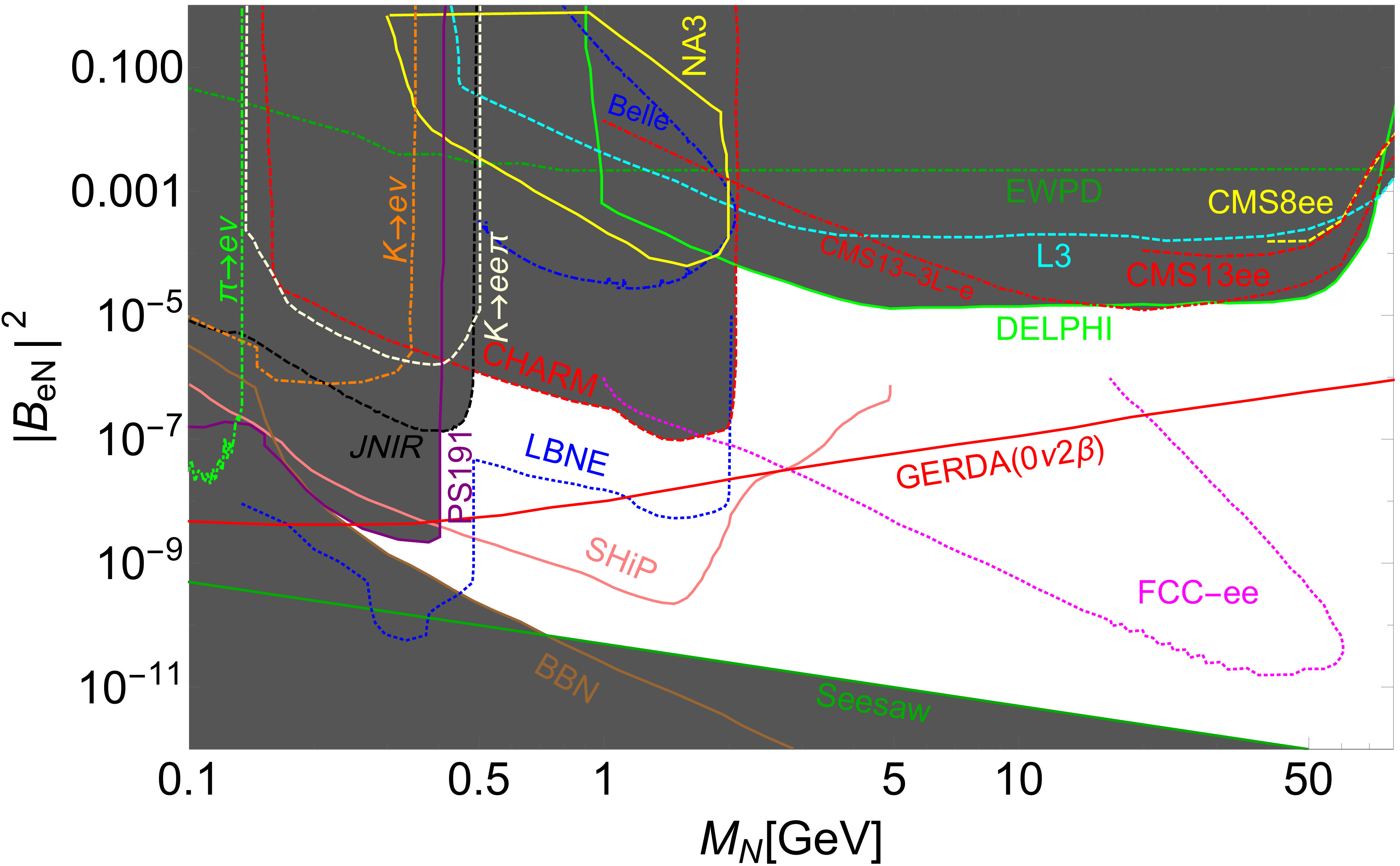}
\caption{Bounds on the light heavy mixing angles for $M_N=0.1$ GeV$-$~$80$ GeV for electron flavor $(|B_{e N}|^2)$.}
\label{fig:mix1}
\end{center}
\end{figure}

\begin{figure}[H]
\begin{center}
\includegraphics[scale=0.3]{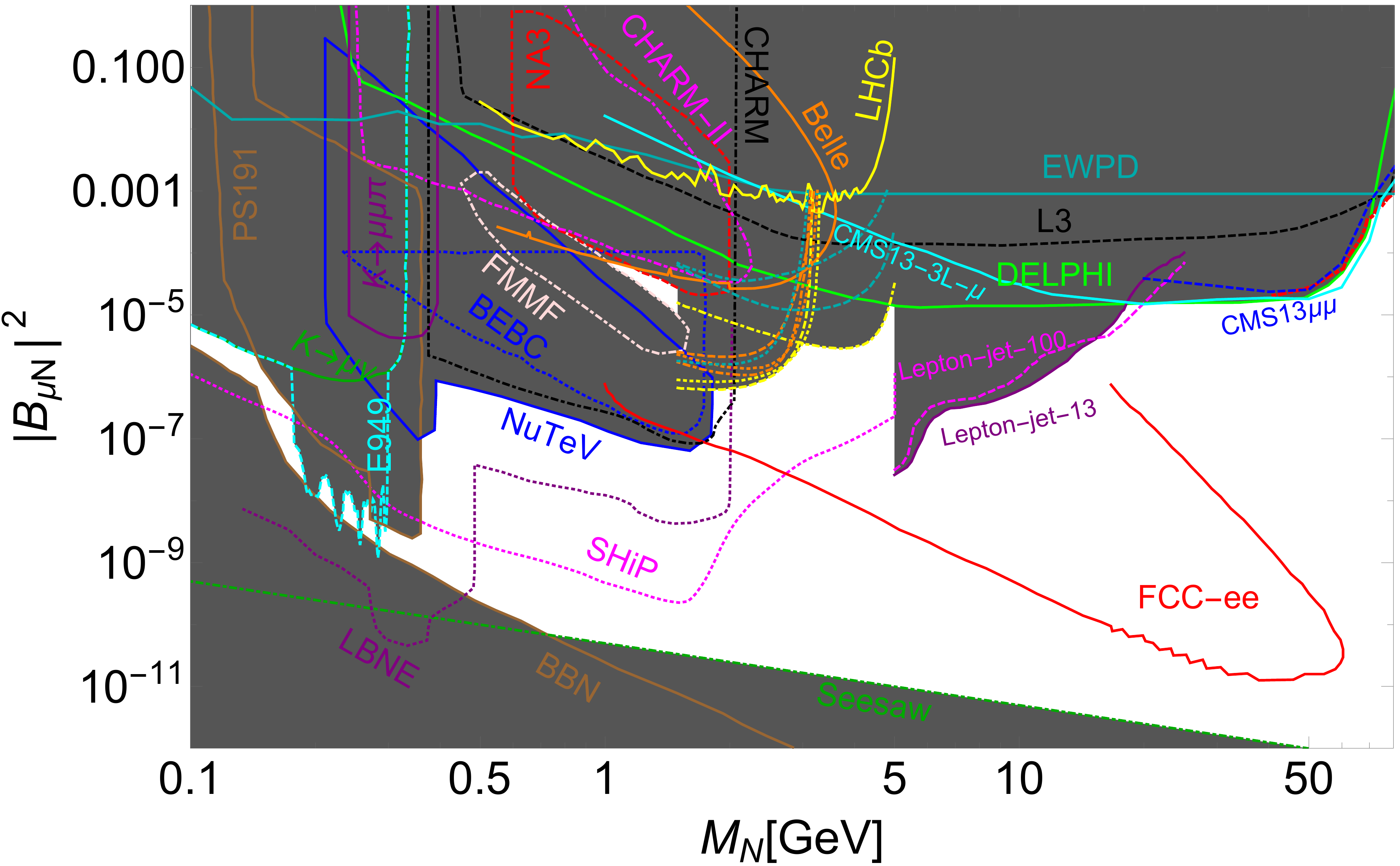}
\caption{Bounds on the light heavy mixing angles for $M_N=0.1$ GeV$-$~$80$ GeV for muon flavor $(|B_{\mu N}|^2)$.
The LHCb (with a detector length of $L = 2.3$ m, and the expected total number of produced $B$ meson pairs $N_B = 4.8 \times 10^{12}$) limits from $B \to D^{*} \mu \mu \pi$, $B\to \mu\mu \pi$ and $B \to D\mu \mu \pi$ are shown by the yellow dashed, yellow dot-dashed and yellow dotted lines respectively. The Belle-II ( effective detector length of $L= 1$ m, and the expected total number of produced $B$ meson pairs $N_B = 5 \times 10^{10}$ ) limits obtained from $B \to \mu\mu\pi, B \to \mu\mu e\nu, B \to D\mu\mu\pi, B \to D^\ast \mu\mu\pi, B \to D \mu\mu e\nu$ and $B \to D^{\ast} \mu\mu e\nu$ are represented by darker cyan dot-dashed, darker cyan dashed, orange dashed, cyan dotted, orange dotted and orange dot-dashed lines respectively \cite{Cvetic:2017vwl, Cvetic:2016fbv} for $1.75$ GeV $< M_N < 5.0$ GeV.}
\label{fig:mix2}
\end{center}
\end{figure}

\begin{figure}[H]
\begin{center}
\includegraphics[scale=0.3]{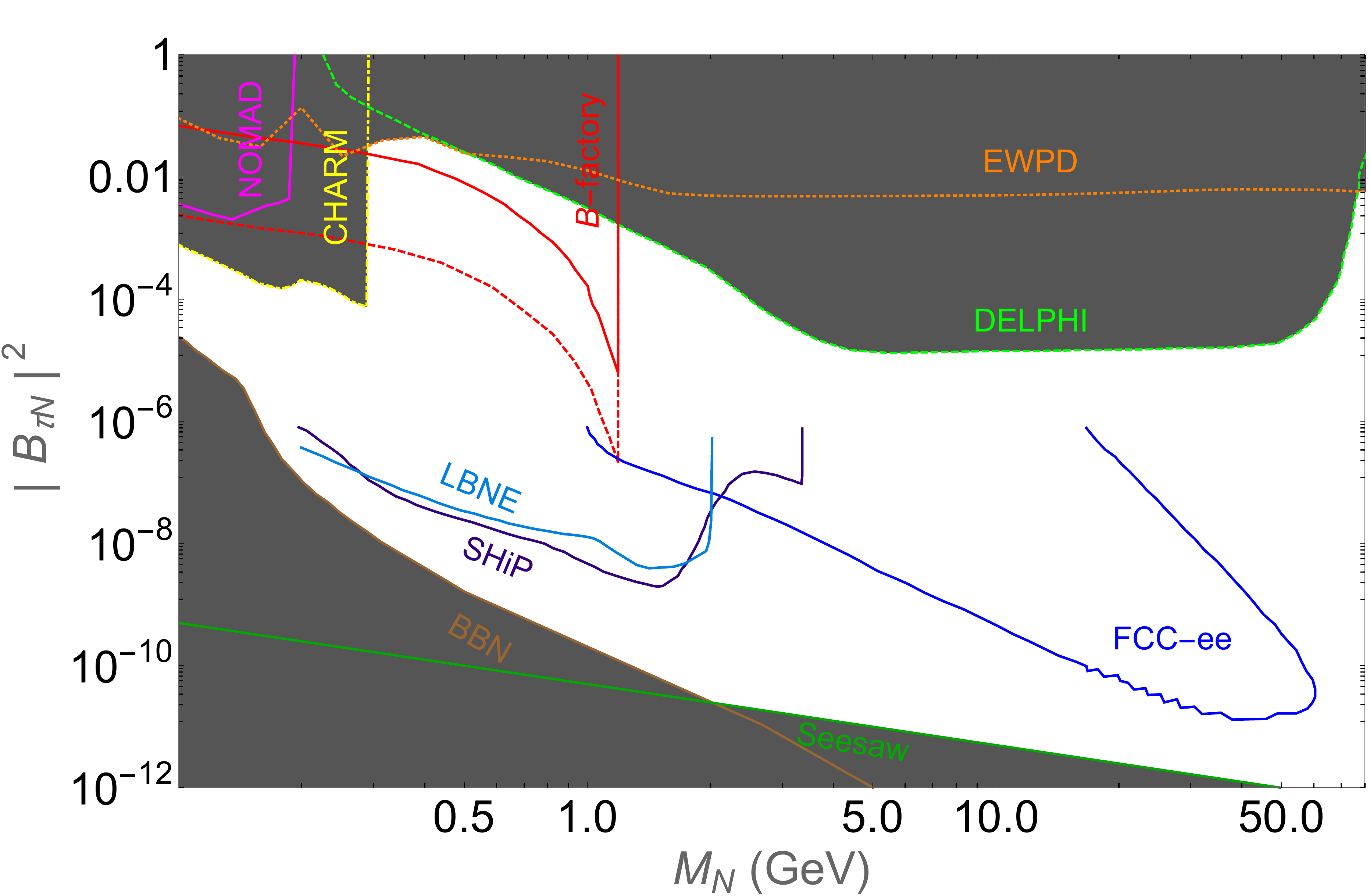}
\caption{Bounds on the light heavy mixing angles for $M_N=0.1$ GeV$-$~$80$ GeV for tau flavor $(|B_{\tau N}|^2)$.}
\label{fig:mix3}
\end{center}
\end{figure}

From the conventional seesaw mechanism we can write 
\bea
\nonumber
B_{\ell N} \sim \sqrt{\frac{m_\nu}{M_N}} 
\eea
and for $m_\nu \leq 0.1$ eV
\bea
B_{\ell N}^2 \leq 10^{-12}\frac{100 GeV}{M_N}
\label{mix}
\eea
which shows the scale of mixing between the light-heavy neutrinos. The limits obtained from Eq.~\ref{mix} have been labeled as `Seesaw' in Figs.~\ref{fig:mix1}- \ref{fig:mix3} respectively.

\begin{figure}[H]
\begin{center}
\includegraphics[scale=0.3]{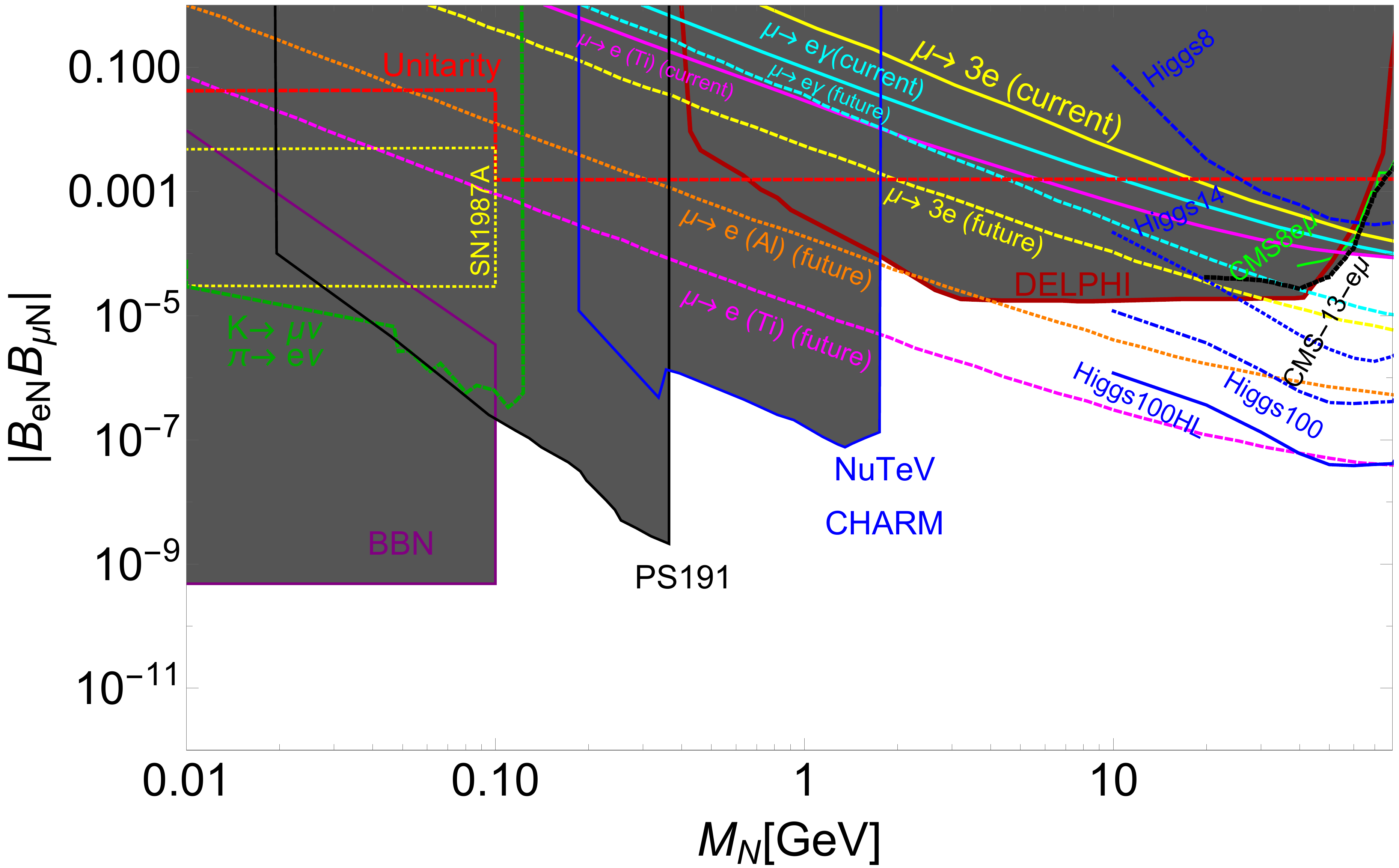}
\caption{Bounds on the light heavy mixing angles $(|B_{eN} B_{\mu N}|)$ for the RHNs with $M_N=0.01$ GeV$-$~$80$ GeV.}
\label{fig:mix4}
\end{center}
\end{figure}

The upper limits on the light-heavy mixing angles have been derived from the cosmological bounds on the RHN lifetimes from the Big Bang Nucleosynthesis (BBN) \cite{Gorbunov:2007ak, Boyarsky:2009ix, Ruchayskiy:2012si}. In the Figs.~\ref{fig:mix1}- \ref{fig:mix3} we plot the upper bounds from different experimental (and prospective) studies on $|B_{e N}|^2$ and $|B_{\mu N}|^2$. In Fig.~\ref{fig:mix4} we plot the upper bounds on $|B_{eN} B_{\mu N}|$. A comprehensive discussion regarding the upper bounds on the mixing angles are also given in \cite{Atre:2009rg, Alonso:2012ji, Deppisch:2015qwa}. 


In Fig.~\ref{fig:mix1} we show the upper bounds on $|B_{eN}|^2$ between the RHN and the light electron neutrino $(\nu_e)$. The heavy Majorana RHNs contribute to the $0\nu2\beta$ decay \cite{Bilenky:2012qi, Bilenky:2014uka} amplitude through the mixing with the light SM neutrinos. The contribution is generated from the exchange of the RHN which depends on the square of the mixing $(|B_{e N_\alpha}|^2)$ between the light weak eigenstate $(\nu_e)$ and a heavy eigenstate $(N_{\alpha R})$, $M_N$ and the Nuclear Matrix Element (NME) due to the RHN exchange \cite{Deppisch:2015qwa, Faessler:2014kka}. The combined search result of $0\nu2\beta$ half life from the GERDA and Heidelberg-Moscow experiment \cite{Agostini:2013mzu} at the $90\%$ C. L. can estimate the upper limit on $|B_{e N}|^2$ with respect to $M_N$. The estimated limits have been expressed as `GERDA' using the red solid curve in Fig.~\ref{fig:mix1}. Such bounds can be less constrained using the effect of cancellation of terms in the expressions of the $0\nu2\beta$ in \cite{Deppisch:2015qwa} followed by \cite{Pascoli:2013fiz} using different elements and their isotopes in the presence of the Majorana phases. In this context we also mention that the generation of the Majorana mass term for the neutrinos from higher dimensional operators has been studied in Ref. \cite{Helo:2015fba} where $0\nu2\beta$ can be observed at the tree level and loop level.


In Figs.~\ref{fig:mix1}- \ref{fig:mix3} the `BBN' in brown represents RHN lifetime greater than one second which is not preferred by the BBN results \cite{Gorbunov:2007ak, Boyarsky:2009ix, Ruchayskiy:2012si}. The exclusion region is shown in gray.
The limits labeled by `Seesaw'  in green represent the scale of mixing from the canonical seesaw mechanism. In both of the cases, the exclusion regions are marked in gray. Such results of BBN and Seesaw may change in the presence of more than two generations 
of RHNs \cite{Drewes:2015iva}. 


The mixing between the light neutrinos and the RHNs can be probed through the weak decays of the mesons and the heavy leptons. Charged mesons can manifest $2$-body decay modes into the RHN in association with a SM charged lepton $(\ell)$. In this case the corresponding branching ratio is proportional to $|B_{\ell N}|^2$. In the meson's rest frame due to the presence of $|B_{\ell N}|^2$ a second monochromatic line in the lepton energy spectrum is expected at the lepton energy apart from the peak due to the active neutrino $(\nu_{i})$. For the RHNs heavier than $\ell$, the helicity suppression factor in the leptonic decay is diluted by $\Big(\frac{M_N}{m_\ell}\Big)^2$ where $m_\ell$ is the mass of $\ell$ \cite{Shrock:1980vy, Shrock:1980ct, Lello:2012gi}.
This is why the sensitivity on $|B_{\ell N}|^2$ increases with $M_N$ with a relevant phase space. Such process is called peak search in meson decay \cite{Deppisch:2015qwa} and schematized in Tab.~\ref{Mes-bary}. 
\begin{table}[t]
\begin{center}
\begin{tabular}{c|c|c}
\hline
\hline
      Experiment type & Decays& Comments\\
\hline
\hline
&$\pi \to e N$ \cite{Azuelos:1986eg, DeLeenerRosier:1991ic, Britton:1992pg, Britton:1992xv, PIENU:2011aa}&The $90\%$ C. L. limits \\
Meson decay&$K \to e N$\cite{K2eN}& for $e~(\mu)$ on $|B_{e N}|^2~(|B_{\mu N}|^2)$   \\
&$K\to \mu N$\cite{K2eN, Asano:1981he,Hayano:1982wu, Kusenko:2004qc, Artamonov:2014urb}& shown in Fig.~\ref{fig:mix1} (Fig.~\ref{fig:mix2}) as `Belle'. \\
\hline
Heavier meson/&$B\to X \ell N, N\to \ell \pi$ \cite{Liventsev:2013zz}&$B\overline{B}$ pair coming from\\
baryon decay\cite{Johnson:1997cj, Ramazanov:2008ph, Gninenko:2009yf}&& $\Upsilon (4s)$ resonance which\\
&&can put $90\%$ C. L. limits on $|B_{e N}|^2$ \\
&&and $|B_{\mu N}|^2$ for the RHN between\\
&& $500$ MeV$\leq M_N \leq 5$ GeV \\
&&are shown as `Belle'.\\
\hline
\end{tabular}
\end{center}
\caption{ Peak search experiments to study the constraints on $|B_{\ell N}|^2$.$\pi\to \mu N$ \cite{Abela:1981nf, Minehart:1981fv, Daum:1987bg, Bryman:1996xd, Assamagan:1998vy} limit is derived between $1$ MeV $\leq M_N \leq 30$ MeV and has been omitted from Fig.~\ref{fig:mix2} due to the choice of $M_N$.
}
\label{Mes-bary}
\end{table}
The $3$-body decay of $\mu$ can set limits on the mixing angles between the light-heavy neutrinos. In this case the Michel spectrum can be distorted by the presence of the a RHN \cite{Shrock:1980ct}.

The unstable RHNs can decay into visible particles through the light-heavy mixings. Therefore the corresponding decay rates are proportional to $|B_{\ell N}|^2$. 
The RHNs can be produced at the semi-leptonic meson decays if kinematically allowed. Once produced from the meson decay, the RHNs decay into visible products like SM leptons, $\pi$ and $K$.
Such visible products can be searched from the beam dump experiments. In this case the detector is placed some distance away from the point of production. We have shown the $90\%$ C. L. limits of such 
experiments in Figs.~\ref{fig:mix1} and \ref{fig:mix2}. The contours are shown in the names of the corresponding experiments, such as PS191\cite{Bernardi:1987ek}, NA3 \cite{Badier:1985wg}, CHARM \cite{Bergsma:1985is,Vilain:1994vg, Orloff:2002de}, IHEP-JINR \cite{Baranov:1992vq}, BEBC\cite{CooperSarkar:1985nh}, FMMF\cite{Gallas:1994xp}, NuTeV \cite{Vaitaitis:1999wq} and NOMAD \cite{Astier:2001ck}. Corresponding limits from NOMAD and CHARM are labeled in Fig.~\ref{fig:mix3}.
The PS191\cite{Bernardi:1987ek} and CHARM \cite{Bergsma:1985is} limits are shown assuming that the RHNs interact only via Eq.~\ref{CC}. The inclusion of Eq.~\ref{NC} needs a new estimation of the bounds in the context of $\nu$SM which makes the bound twice stronger \cite{Ruchayskiy:2011aa}. The excluded regions form the current experimental data are shaded in gray. 

Long Baseline Neutrino Experiment (LBNE) (currently known as Deep Underground Neutrino Experiment (DUNE)) may probe smaller values of the mixing with a near detector \cite{Adams:2013qkq}. The expected exclusion contour has been shown as `LBNE' in each of Figs.~\ref{fig:mix1}- \ref{fig:mix3} considering the normal mass ordering of the light neutrinos.


The Lepton Number Violating (LNV) decays of the mesons $(X)$ are forbidden in the SM. However, SM extended by the RHNs allows the LNV rare meson decays $(X_i \to \ell^{\pm} N, N \to \ell^{\pm} X_j^{\mp})$.
Such models will be interesting in the experiments like  CLEO, Belle, BABAR and LHCb. The strongest bound is coming from the $K\to \ell^{+}\ell^{+} \pi^{-}$ mode 
\cite{Atre:2009rg} as shown in Fig.~\ref{fig:mix1} for $\ell=e$ and in Fig.~\ref{fig:mix2} for $\ell=\mu$  assuming a $10$ meter detector size. The limits from the $D$ and $B$ meson decays \cite{Castro:2013jsn, Yuan:2013yba, Wang:2014lda} are weaker than the current LHCb bounds \cite{Aaij:2014aba}. The revised LHCb bounds are shown in Fig.~\ref{fig:mix2} using $B^-\to \pi^+ \mu^-\mu^-$ with $3$ fb$^{-1}$ of integrated luminosity at the LHC Run-I with $\sqrt{s}=7$ and 8 TeV.
The experimentally excluded parameter spaces are shaded in gray in Fig.~\ref{fig:mix2} from \cite{Shuve:2016muy} which has been obtained using proper decay width formulae for the RHNs having mass between $0.5$ GeV $\leq M_N \leq 5$ GeV and considering $|B_{e N}|=|B_{\tau N}|=0$. The revised limits for $|B_{\mu N}|^2$ are obtained from Belle \cite{Liventsev:2013zz} in Fig.~\ref{fig:mix2} are also shown in \cite{Shuve:2016muy}. The revised Belle limits on $|B_{eN}|^2$ in Fig.~\ref{fig:mix1} are taken from \cite{Liventsev:2013zz}, see the erratum of \cite{Liventsev:2013zz} for more information. In this context, we also mention that LHCb upgrade with a detector length of $L = 2.3$ m, and the expected total number of produced $B$ meson pairs $N_{B} = 4.8 \times 10^{12}$
may reach for $|B_{\mu N}|^2$ the upper bounds of about $6 \times 10^{-6}$ for $1.7$ GeV $< M_N < 2.5$ GeV, coming from the rare lepton number violating decays of the $B$ meson such as $B \to D^{*} \mu \mu \pi$ \cite{Cvetic:2017vwl, Cvetic:2016fbv}. The yellow dashed line represents the limits obtained from $B \to D^{*} \mu \mu \pi$ in Fig.~\ref{fig:mix2}. The corresponding limits from $B\to \mu\mu \pi$ are shown by the  yellow dot-dashed line in Fig.~\ref{fig:mix2} whereas the limits coming from $B \to D\mu \mu \pi$ are represented by the yellow dotted line. The future Belle-II sensitivity limits on $|B_{\mu N}|^2$ from the lepton number violating decays of $B$ meson are also shown in Fig.~\ref{fig:mix2} from \cite{Cvetic:2017vwl}. The corresponding decay modes are $B \to D^{\ast} \mu^{\pm} N, N \to \mu^{\pm} X^{\mp}$ at Belle-II, where $X^{\mp} = \pi^{\mp}$ or $X^{\mp} = e^{\mp}\nu_e$;  $B\to D\mu^\pm N \to D \mu^\pm \mu^\pm X^\mp$; $B^\pm \to \mu^\pm N \to \mu^{\pm}\mu^{\pm}X^{\mp}$. We compare all these limits of the prospective Belle-II sensitivity in Fig.~\ref{fig:mix2} considering an effective detector length of $L= 1$ m, and the expected total number of produced $B$ meson pairs $N_B = 5 \times 10^{10}$. The limits obtained from $B \to \mu\mu\pi, B \to \mu\mu e\nu, B \to D\mu\mu\pi, B \to D^\ast \mu\mu\pi, B \to D \mu\mu e\nu$ and $B \to D^{\ast} \mu\mu e\nu$ are represented by darker cyan dot-dashed, darker cyan dashed, orange dashed, cyan dotted, orange dotted and orange dot-dashed lines respectively. The Belle-II can improve the sensitivity down to $2\times 10^{-6}$ on $|B_{\mu N}|^2$.

The decay length of the RHN can be shorter than the size of the detector. In that case the number of the signal events will be suppressed by the acceptance factor $P_N$. On the other hand, if the decay length is larger than the detector size, the RHN will decay outside the detector and its detection will not be possible. In that case the signal events will be less. The limitation can be fixed by increasing the flux of the initial hadrons in a proposed fixed target experiment like SHiP \cite{Anelli:2015pba}. In this case a high intensity proton beam is planned to be used at the CERN SPS. In this experiment the beam dump technique can help to get rid of the huge SM background to propagate the RHNs freely. Such an experiment can probe $|B_{\ell N}|^2$ below $10^{-9}$ \cite{Alekhin:2015byh}. The corresponding projection contours are shown as `SHiP' in Figs.~\ref{fig:mix1}- \ref{fig:mix3}.
If no such decays are detected, these limits may soon significantly improve the present limits which are in black in Fig.~\ref{fig:mix2}, where we note that the restrictive bounds from BEBC and NuTeV exist only up to $M_N \approx 1.7$ GeV.

$\tau$ can decay into $NX$ where $X$ stands for the hadrons. The mass and energy of $X$ can be reconstructed at a high precision once it is hadronized into either of $\pi$ or $K$. 
Using the $\tau$ decay mode of $\tau \to N \pi^-\pi^+\pi^-$, the limits on $|V_{\tau N}|^2$ have been shown in Fig.~\ref{fig:mix3}. The limits have been labeled as B-factory. 
The optimistic projected limits have been produced at the $90\%$ C. L. of nearly ten million of $\tau$ decays.
The results from the peak search experiments are very strong because of the usage of the kinematic features and minimal assumptions regarding the decays of the RHNs as they are assumed to be produced on-shell.
These limits are valid for the RHNs irrespective of the Majorana or Dirac type.


The RHNs lighter than $Z$ boson can be produced from the $Z$ decay following the NC interaction given in Eq.~\ref{NC}. \cite{Dittmar:1989yg, Das:2017pvt}.
The $95\%$ C. L. limits on $|B_{\ell N}|^2$ were obtained from L3 \cite{Adriani:1992pq} and DELPHI \cite{Abreu:1996pa} analyzing the LEP data. These results were shown as the `L3' and `DELPHI' 
curves in Figs.~\ref{fig:mix1}- \ref{fig:mix3}. respectively.

A future search in the proposed FCC-ee experiment can improve the sensitivity down to $|B_{\ell N}|^2 \sim 10^{-12}$ for all the neutrino flavors with RHNs between $10$ GeV $\leq M_N \leq 80$ GeV \cite{Blondel:2014bra,Abada:2014cca}.
The limits on the individual mixings were estimated assuming the normal mass hierarchy of the light neutrinos \cite{Blondel:2014bra} assuming a $10^{12}$ Z decays between $10-100$ cm from the point of interaction. 
The sensitivity can be made better if the number of $Z$ bosons (the range of decay length) are (is) increased which can reach at the theoretical expectation labeled as `Seesaw' in the corresponding figures Figs.~\ref{fig:mix1}- \ref{fig:mix3}.


The RHNs can affect the electroweak precision searches due to its mixing with the SM light neutrinos. The observation takes place through the invisible decay of the $Z$ boson \cite{Nardi:1994iv,Nardi:1994nw,Abada:2012mc}. $|B_{\ell N}|^2$ also affects the unitarity of the lepton mixing matrix \cite{Antusch:2006vwa,Abada:2007ux,Antusch:2014woa}. The presence of the mixing affects the lepton universality in the leptonic and semi-leptonic decays of the pseudoscalar mesons \cite{Abada:2012mc, Antusch:2014woa, Abada:2013aba, Asaka:2014kia}. Stringent constraints are derived on $ |B_{\ell N}|^2$ using the global fits to the Electroweak Precision Data (EWPD) at the $90\%$ C. L \cite{delAguila:2008pw,Akhmedov:2013hec,deBlas:2013gla, Basso:2013jka,Antusch:2015mia}. The results are labeled as `EWPD' in Figs.~\ref{fig:mix1}- \ref{fig:mix3}. For $M_N > M_Z$, these limits are independent of RHN mass, however, a mass dependence is observed for $M_N < M_Z$. 

Direct collider searches for the Majorana type RHNs are performed by the LHC at the CMS \cite{Khachatryan:2015gha,Khachatryan:2016olu} for $M_N \leq 80$ GeV. These mixing bounds from the CMS at the $\sqrt{s}=8$ TeV are calculated for the same sign dilepton plus dijet final state which are labeled as `CMS$8$ee' in Fig.~\ref{fig:mix1} for the $e$ flavor using yellow dashed curve and `CMS$8\mu\mu$' in Fig.~\ref{fig:mix2} for $\mu$ flavor using the red dot-dashed curve respectively. Limits for the RHNs using the lepton-jet technique for the $\mu$ flavor are shown in Fig.~\ref{fig:mix2} between $5$ GeV $\leq M_N \leq 25$ GeV  at the $13$ TeV LHC and a prospective $100$ TeV proton proton collider \cite{Dube:2017jgo}. The exclusion limits at the $13$ TeV LHC ($100$ TeV) are represented by `Lepton-jet-13' (`Lepton-jet-100'). The prospective bounds obtained at a future $100$ TeV proton proton collider are better than those obtained at the $13$ TeV LHC  for $M_N > 15$ GeV using the lepton-jet \cite{Dube:2017jgo}. The current limits are obtained from the CMS at the $13$ TeV using the same sign dilepton \cite{CMS:2018szz} and trilepton \cite{Sirunyan:2018mtv} modes. The limits from the same sign dileption on $|B_{\ell N}|^2$ using the $e$ flavor is marked as `CMS $13$ee' in Fig.~\ref{fig:mix1}. This limit is shown by the  red dashed curve and it is weaker than the DELPHI limits roughly by one order of magnitude. The `CMS $8$ee' results match with `CMS $13$ee' for $M_N \geq 60.2$ GeV. The same sign dilepton limit from the $\mu$ flavor is marked as `CMS $13\mu\mu$' and represented by blue dashed curve in Fig.~\ref{fig:mix2}. The limits are weaker than the DELPHI results for $M_N \leq 72$ GeV. The `CMS $8\mu\mu$' results are stronger than the `DEPLHI' limits for $M_N \geq 70$ GeV and better than the `CMS $13\mu\mu$' results. 
The $13~(8)$ TeV CMS limits are calculated at the $35.9~(19.7)$ fb$^{-1}$ luminosity. The $8$ TeV CMS results for $e$ flavor are estimated at the $19.7$ fb$^{-1}$ luminosity. The ATLAS results at $8$ TeV with $20.3$ fb$^{-1}$ luminosity are studied in \cite{Aad:2015xaa} considering RHNs heavier than the SM $Z$ boson $(100$ GeV $\leq M_N\leq 500$ GeV$)$, however, that mass range is beyond the scope of this paper. The trilepton bounds for the Majorana neutrinos are taken from the CMS analysis \cite{Sirunyan:2018mtv} at the $13$ TeV LHC with a luminosity of $35.9$ fb$^{-1}$. The limits on $|B_{eN}|^2$ are shown in Fig.~\ref{fig:mix1} by `CMS$13$-$3$L-$e$' with red dot-dashed curve. In this trilepton case the first two leptons associated with RHN are of the $e$ flavor. This result is better than those of the `DELPHI' results between $15.6$ GeV $\leq M_N \leq 24.0$ GeV and `L3' results are better than this for $M_N \geq 72.8$ GeV. The trilepton case, where the first two leptons associated with the RHN are of the $\mu$ flavor, puts bounds on $|B_{\mu N}|^2$ in Fig.~\ref{fig:mix2}. It is designated by solid cyan curve and `CMS$13$-$3$L-$\mu$'. This is better than the `DELPHI' limit for $M_N \geq 48$ GeV, however, `CMS$8\mu\mu$' is better than this limit for $M_N \geq 73$ GeV. We can compare the dilepton and trilepton limits because both are probing the Majorana nature of the RHN through the LNV mode. 


For the seesaw scenario $Y_D$ is a $3\times 3$ non-diagonal matrix. We consider $m_N$ as a $3\times 3$ diagonal matrix so that a flavor non-diagonal scenario can be studied as done in the Ref.\cite{Das:2017nvm} for a minimal scenario. We briefly summarize the current experimental bounds on the mixing angles in Fig.~\ref{fig:mix4} coming from the LFV processes such as $\mu \to e\gamma$ using \cite{Atre:2009rg,Alonso:2012ji,Deppisch:2015qwa} considering RHNs between $0.1$ GeV $< M_N < 80$ GeV. The non-unitarity of the lepton mixing matrix put constraints on the mixing matrix according to \cite{Antusch:2006vwa, Antusch:2008tz, Antusch:2014woa}. Using the current $\mu \to e \gamma$ branching ratio from \cite{TheMEG:2016wtm} and following exactly the same procedure mentioned in \cite{Alonso:2012ji} we update the current limit for on the mixing angle $(|B_{eN} B_{\mu N}|)$ for the RHNs from leptonic non-unitarity. The corresponding bounds have been labeled as `Unitarity' in Fig.~\ref{fig:mix4}. The light RHNs can be produced from the meson decay \cite{Hayano:1982wu, Yamazaki:1984sj, Britton:1992xv} in association with the $\ell$. The shaded regions are excluded by the $K$ meson decay searches from PS$191$ experiment \cite{Bernardi:1987ek}, $D$ meson from CHARM \cite{Bergsma:1985is} and NuTeV \cite{Vaitaitis:1999wq}. These results were collected at the $90\%$ C. L. In PS$191$ experiment Dirac type RHNs were studied between $10$ MeV $< M_N < 400$ MeV for the LFV signals. Such bounds can be converted into the Majorana ones just roughly multiplying by a factor of $2$ to obtain the results for $e^{\pm}\mu^{\pm}$ final sates from the Majorana RHNs. In a similar way the limits for the characteristic LFV signal for the Majorana RHNs can be extracted from the CHARM and NuTeV experiments. The non observations of the $e^{\pm}\mu^{\pm}$ final sates from the meson decays set very strong limits for the RHNs between $0.001$ GeV $< M_N <  2.0$ GeV. RHNs lighter than the $Z$ bosons are constrained by the $Z \to N \nu$ searches at DELPHI \cite{Abreu:1996pa} at the $95\%$ C. L. For $M_N <  2$ GeV, the constraints set by DELPHI are weaker than those obtained by the peak search experiments. The SN$1897$A data can put limits on $|B_{eN} B_{\mu N}|$ \cite{Kainulainen:1990bn, Kusenko:2004qc,Mangano:2011ar, Ruchayskiy:2012si}. Cosmological studies also put strong limits on the mixing angles as discussed in \cite{Drewes:2015iva,Drewes:2012ma, Canetti:2012kh, Drewes:2013gca, Canetti:2014dka, Adhikari:2016bei, Drewes:2016jae, Antusch:2017pkq, Antusch:2017hhu, Antusch:2016ejd, Antusch:2016vyf,Abada:2017jjx}. Current (future) limits from $\mu \to 3e$, $\mu \to e$ can be found from \cite{Alonso:2012ji} and \cite{Bellgardt:1987du, Berger:2011xj} and are represented by the solid (dashed) curves.

Bounds from the SM Higgs $(H)$ decay into RHNs are also interesting in this context when $H \to N \nu, N \to \ell \ell \nu $ can be produced at the colliders for the RHNs between $10$ GeV $< M_N < 80$ GeV \cite{BhupalDev:2012zg, Das:2017zjc, Das:2017rsu}. 
It shows that the LHC Higgs searches at the $8$ TeV (Higgs8) and $14$ TeV (Higgs14) can put constraints on the mixing angles if $H$ decays into RHNs through the Yukawa coupling in Fig.~\ref{fig:mix4}. The prospective $100$ TeV proton proton collider at high luminosity $(30000)$ fb $^{-1}$ could probe the mixings better than the current limits for RHNs with $M_N > 40$ GeV at the high luminosity (Higgs$100$HL). The $100$ TeV results at the  $3000$ fb $^{-1}$ luminosity will have lower sensitivity (Higgs$100$). The detailed analyses of these results are given in \cite{Das:2017zjc}. Direct experimental search for the Majorana RHNs at the LHC is performed by the CMS with $13$ TeV energy and $35.9$ fb$^{-1}$ luminosity for the LFV same sign dilepton with $e^{\pm}\mu^{\pm}$ in the final state \cite{CMS:2018szz}. This limits are denoted by `CMS$-13-e\mu$' and represented by a black dotted curve from $M_N \geq 20$ GeV in Fig.~\ref{fig:mix4}. This result is stronger than the LFV `DELPHI' limits for $M_N \geq 44$ GeV and weaker than $\mu \to e (Ti)$
for $M_N \geq 57$ GeV.

\begin{figure}[H]
\begin{center}
\includegraphics[scale=0.3]{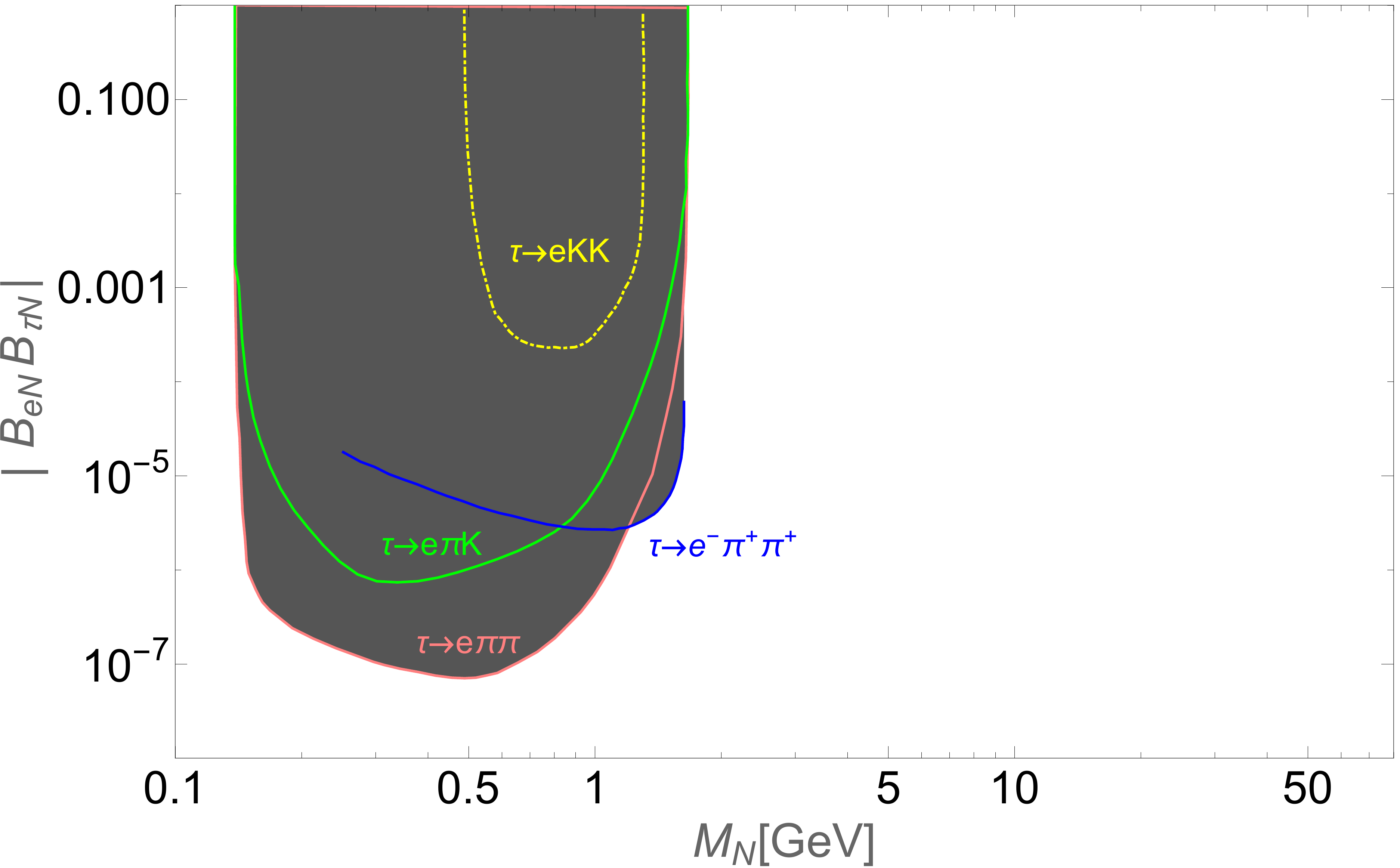}
\caption{Bounds on the light heavy mixing angles $(|B_{eN} B_{\tau N}|)$ for the RHNs with $M_N=0.1$ GeV$-$~$80$ GeV.}
\label{fig:mix5}
\end{center}
\end{figure}

\begin{figure}[H]
\begin{center}
\includegraphics[scale=0.3]{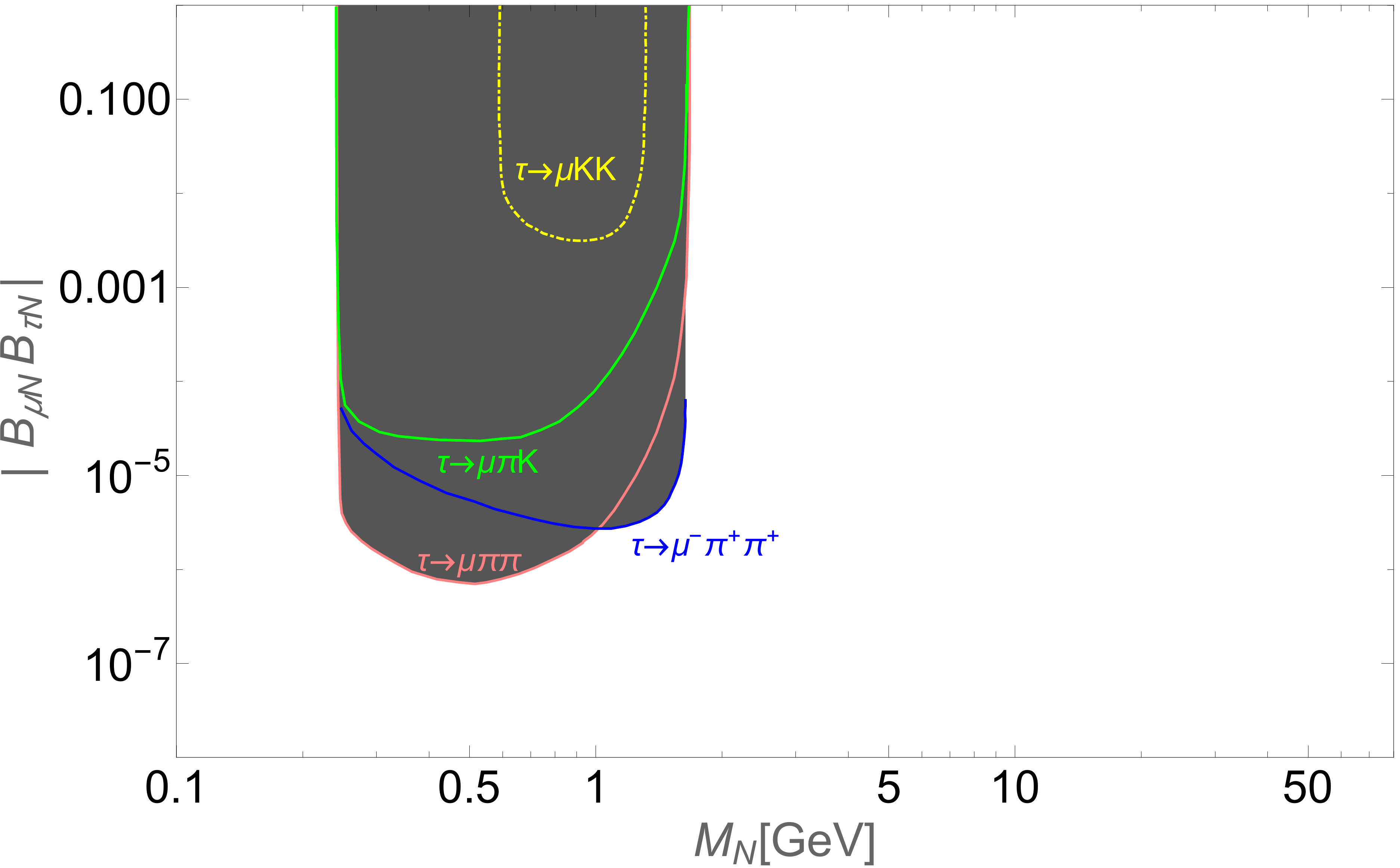}
\caption{Bounds on the light heavy mixing angles $(|B_{\mu N} B_{\tau N}|)$ for the RHNs with $M_N=0.1$ GeV$-$~$80$ GeV.}
\label{fig:mix6}
\end{center}
\end{figure}

The mixing limits from the $\tau$ decay into hadrons in association with $e$ and $\mu$ are shown in Figs.~\ref{fig:mix5} and \ref{fig:mix6} respectively from BABAR \cite{Atre:2009rg,Aubert:2005tp}. The limits on $\tau\to e \pi\pi$, $\tau \to e \pi K$, $\tau \to e K K$ and $\tau \to e^-\pi^+\pi^+$ are shown in Fig.~\ref{fig:mix5} and those with $\mu$ are shown in Fig.~\ref{fig:mix6}. A recent study \cite{Zamora-Saa:2016ito} of the LNV $\tau$ decays via intermediate on-shell Majorana neutrinos set bound on the RHNs having mass within $0.2$ GeV $\leq M_N \leq 1.68$ GeV. The limits are stringent for the RHNs having mass, $M_N > 1$ GeV and goes down to $3.082\times 10^{-6}$ at $M_N =1.245$ GeV for the $e$ and $2.795\times10^{-6}$ at $M_N=1.042 \times 10^{-6}$ for the $\mu$ respectively.


\section{Production of the RHN}
\label{s3}

As we stated above, we are interested in studying the Lepton Number Violating (LNV) and Lepton Number Conserving (LNC) in $W^{\pm}$ boson\footnote{In Appendix \ref{app4} we extend the formalism to top quark decays.} decays mediated by at least two heavy Majorana neutrinos. 
\subsection{$W^{\pm}$ Boson Decays}
\begin{figure}[H]
\centering
\includegraphics[scale = 0.66]{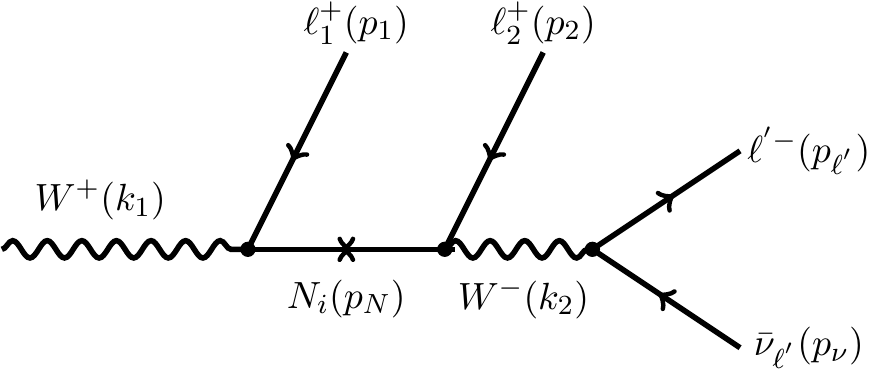}\hspace{1 cm}
\includegraphics[scale = 0.66]{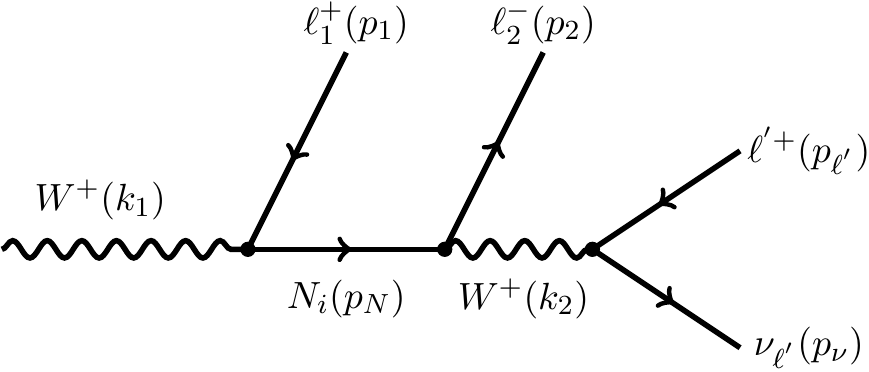}
\caption{ Feynman diagrams for the LNV process $W^+\rightarrow \ell_1^+ \ell_2^+ \ell^{' -} \bar{\nu}_{\ell^{'}}$ (left panel) and LNC process $W^+\rightarrow \ell_1^+ \ell_2^- \ell^{' +} {\nu}_{\ell^{'}}$ (right panel).}
\label{fig:Wdecays}
\end{figure}

The relevant Feynman diagrams for these processes are presented in Fig.~\ref{fig:Wdecays} 

In order to obtain the decay amplitude we have to write down all the relevant quantities. The neutrino flavor state is
\begin{equation}
\label{fstat}
\nu_{\ell}=\sum_{i=1}^3 B_{\ell i}\nu_i + \sum_{j=1}^n B_{\ell N_j} N_j \ ,
\end{equation}
where $B_{\ell N_j}$ are the the heavy-light mixing elements of the PMNS matrix, which can be written as 
\begin{equation}
\label{PMNS}
 B_{\ell N_j} =  |B_{\ell N_j}| e^{i \phi_{\ell N_j}} \ .
\end{equation}

\begin{figure}[H]
\centering
\includegraphics[scale = 0.65]{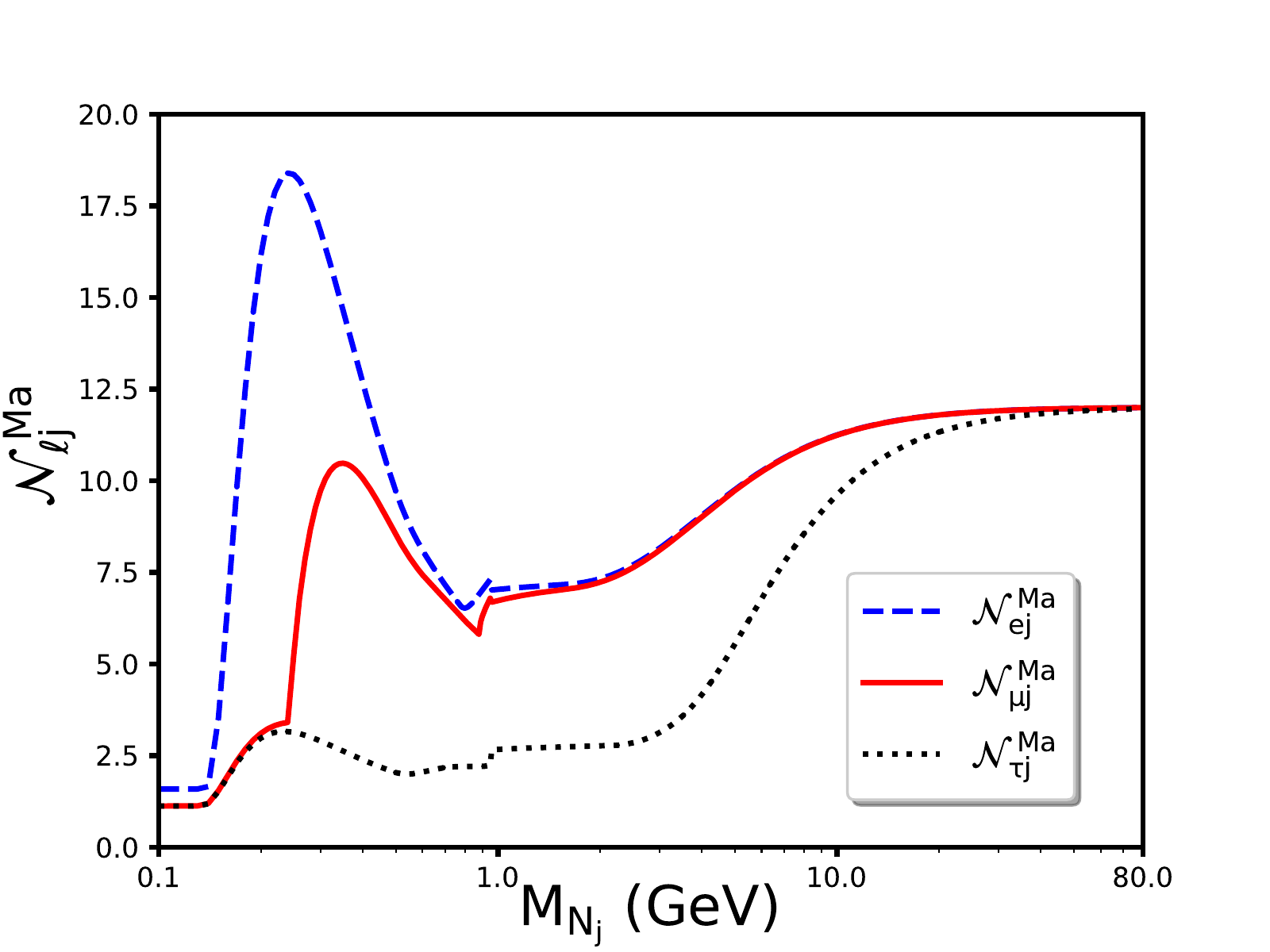}
\caption{Effective mixing coefficients ${\cal N}_{\ell j}^{\rm Ma}$ for Majorana neutrinos.}
\label{fig:efcoef}
\end{figure}

From now on, we will use (${\rm LV}$) and (${\rm LC}$) instead (LNV) and (LNC) in the equations, in order to have a shorter notation. The averaged square amplitudes for lepton number violating and lepton number conserving decay processes, involving $n$ sterile neutrinos, are
\begin{align}
\nonumber \overline{|\mathcal{M}_{W^{+}}^{\rm LV}|^2_{ij}}&=\frac{4 \sqrt{2}}{3}\, G_F^3 \ B^{*}_{l_1 N_i} B^{*}_{l_2 N_i} B_{l_1 N_j} B_{l_2 N_j} M_{N_i} M_{N_j} \times  \\ 
 & \bar{G}_{N_i} \bar{G}^{*}_{N_j} \Bigg( \frac{1}{(k_2^2-M_W^2)^2-\Gamma_W^2 M_W^2} \Bigg )  f^{\rm LV}_{\rm kin}(k_1,p_1,p_2,p_{\ell^{'}},p_{\nu}) \ ,
 \label{ampLNV}
\end{align}
\begin{align}
\nonumber \overline{|\mathcal{M}_{W^{+}}^{\rm LC}|^2_{ij}}&=\frac{4 \sqrt{2}}{3}\, G_F^3 \ B^{*}_{l_1 N_i} B_{l_2 N_i} B_{l_1 N_j} B^{*}_{l_2 N_j} \times \\
& \bar{G}_{N_i} \bar{G}^{*}_{N_j} \Bigg( \frac{1}{(k_2^2-M_W^2)^2-\Gamma_W^2 M_W^2} \Bigg ) f^{\rm LC}_{\rm kin}(k_1,p_1,p_2,p_{\ell^{'}},p_{\nu}) \ .
\label{ampLNC}
\end{align}
Here $k_2^2={m_{\ell^{'}}}^2+2 (p_{\ell^{'}} \cdot p_{\nu})$, the kinematical functions $f^{\rm X}_{\rm kin}(k_1,p_1,p_2,p_{\ell^{'}},p_{\nu})$ ($X=LV,\ LC$) are given by Eqs.~\eqref{kinLV} and \eqref{kinLC}, respectively, which are presented in Appendix~\ref{app1}. The factors $\bar{G}_{N_i}$ are the heavy neutrino propagators 
\begin{align}
\bar{G}_{N_i}= \frac{1}{p_N^2 -M_{N_i}^2+i \Gamma_{N_i} M_{N i}} \equiv \frac{1}{(k_1 - p_1)^2 -M_{N_i}^2+i \Gamma_{N_i} M_{N i}}
\end{align}
where $\Gamma_{N_i}$ is the total decay width of the intermediate Majorana neutrino, which can be approximated as follows \cite{Cvetic:2014nla}
\begin{equation}
\Gamma_{\rm Ma}(M_{N_i})  \approx  \K_i^{\rm Ma}\ \frac{G_F^2 M_{N_i}^5}{96\pi^3} \, ,
\label{DNwidth}
\end{equation}
where the coefficients $\K_i^{\rm Ma}$ are given by
\begin{equation}
\K_i^{\rm Ma} = {\cal N}_{e i}^{\rm Ma} \; |B_{e N_i}|^2 + {\cal N}_{\mu i}^{\rm Ma} \; |B_{\mu N_i}|^2 + {\cal N}_{\tau i}^{\rm Ma} \; |B_{\tau N_i}|^2
\ .
\label{calK}
\end{equation}
Here, the factors ${\cal N}_{\ell i}^{\rm Ma}$ are the effective mixing coefficients, which account for all possible decay channels of $N_i$ and are presented in Fig.~\ref{fig:efcoef} for our mass range of interest.
We also notice that charge-conjugate processes satisfy the symmetry relations
\begin{equation}
 \overline{|\mathcal{M}_{W^{-}}^{\rm LV}|^2_{ij}}=\Big(\overline{|\mathcal{M}_{W^{+}}^{\rm LV}|^2_{ij}} \Big)^* \quad , \quad  
 \overline{|\mathcal{M}_{W^{-}}^{\rm LC}|^2_{ij}}=\Big(\overline{|\mathcal{M}_{W^{+}}^{\rm LC}|^2_{ij}} \Big)^* \ ,
\end{equation}
where overbar denotes the sum over the final and average over the initial helicities.
The $W^{\pm}$ boson decay widths, neglecting the effects of heavy neutrino oscillations, are  
\begin{subequations}
\label{dw}
\begin{align}
\Gamma(W^{+} \rightarrow \ell_1^{+} \ell_2^{+} {\ell}^{' -} \bar{\nu}_{\ell^{'}}) &\equiv \Gamma_{LV}(W^{+}) = \frac{1}{2 M_{W} (2 \pi)^{8}} \int \ \sum_{i,j=1}^{n} \overline{|\mathcal{M}_{W^+}^{LV}|^2_{ij}} \ d_4 \quad, \\
\Gamma(W^{+} \rightarrow \ell_1^{+} \ell_2^{-} {\ell}^{' +} \nu_{\ell^{'}}) &\equiv \Gamma_{LC}(W^{+}) = \frac{1}{2 M_{W} (2 \pi)^{8}} \int \ \sum_{i,j=1}^{n}  \overline{|\mathcal{M}_{W^+}^{LC}|^2_{ij}} \ d_4 \quad,
\end{align}
\end{subequations}
\begin{subequations}
\label{dw}
\begin{align}
\Gamma(W^{-} \rightarrow \ell_1^{-} \ell_2^{-} {\ell}^{' +} {\nu}_{\ell^{'}}) &\equiv \Gamma_{LV}(W^{-}) = \frac{1}{2 M_{W} (2 \pi)^{8}} \int \ \sum_{i,j=1}^{n} \overline{|\mathcal{M}_{W^-}^{LV}|^2_{ij}} \ d_4 \quad, \\
\Gamma(W^{-} \rightarrow \ell_1^{-} \ell_2^{+} {\ell}^{' -} \bar{\nu}_{\ell^{'}}) &\equiv \Gamma_{LC}(W^{-}) = \frac{1}{2 M_{W} (2 \pi)^{8}} \int \ \sum_{i,j=1}^{n}  \overline{|\mathcal{M}_{W^-}^{LC}|^2_{ij}} \ d_4 \quad.
\end{align}
\end{subequations}
Here $d_4$ denotes the number of states available per unit of energy in the 4-body final state and is given as follows
\begin{equation}
d_4 \equiv \frac{d^3 {\vec p}_1}{2 E_{1}({\vec p}_1)} 
 \frac{d^3 {\vec p}_2}{2 E_{2}({\vec p}_2)}
 \frac{d^3 {\vec p}_{\ell^{'}}}{2 E_{\ell^{'}}({\vec p}_{\ell^{'}})}
 \frac{d^3 {\vec p}_{\nu}}{2 E_{\nu}({\vec p}_{\nu})}
\delta^{(4)} \left( k_1 - p_1 - p_2 - p_{\ell^{'}} - p_{\nu} \right) \ ,
\label{d4}
\end{equation}
From now on we will work in a scenario with two heavy neutrinos in the mass range $M_{N_i} \sim$ $1$-$80$ GeV. Therefore, after the  integrations are implemented, the decay widths are
\begin{align}
\label{DWE}
\nonumber \Gamma_{X}(W^{\pm}) &=|B_{\ell_{1} N_1}^{}|^2 |B_{\ell_{2} N_1}^{}|^2 \frac{\widetilde{\Gamma}_{X}(M_{N_1})}{\Gamma_{\rm Ma}(M_{N_1})}+ |B_{\ell_{1} N_2}^{}|^2 |B_{\ell_{2} N_2}^{}|^2\frac{\widetilde{\Gamma}_{X}(M_{N_2})}{\Gamma_{\rm Ma}(M_{N_2})}\\
&+2\; \frac{\widetilde{\Gamma}_{X}(M_{N_1})}{\Gamma_{\rm Ma}(M_{N})} \; |B_{\ell_1 N_1}^{}| |B_{\ell_1 N_2}^{}| |B_{\ell_2 N_1}^{}| |B_{\ell_2 N_2}^{}| \Big( \cos(\theta_{X}) \delta_{12} \mp \frac{\eta(y)}{y} \sin(\theta_{X})  \Big) \ ,\\
\end{align}
Here $\theta_{LV}$ and $\theta_{LC}$  are the CP violating angles and are given as follow
\begin{subequations}
\begin{align}
\theta_{LV} &\equiv \phi_{\ell_1 N_2} + \phi_{\ell_2 N_2} - \phi_{\ell_1 N_1} - \phi_{\ell_2 N_1}  \ , \\
\theta_{LC} &\equiv \phi_{\ell_1 N_2} + \phi_{\ell_2 N_1} - \phi_{\ell_1 N_1} - \phi_{\ell_2 N_2} \ .
\end{align}
\end{subequations}
The factor $\delta_{12}\equiv$ $\Re \big[ \int  \overline{|\mathcal{M}^{X}_{W^{\pm}}|^2_{12}} \ d_4 \big] \big/ \int  \overline{|\mathcal{M}^{X}_{W^{\pm}}|^2_{11}} \ d_4 $ measures the effect of $N_1 - N_2$ overlap\footnote{$\Re$ stands for the real part.}  and its values are presented in Fig.~\ref{delta_eta} (Appendix~\ref{app2}) as a function of $y\equiv \frac{\Delta_{M_N}}{\Gamma_{\rm Ma}}=\frac{M_{N_2}-M_{N_1}}{\Gamma_{\rm Ma}}$. The factor $\eta(y)$ parametrizes the deviations from the Narrow Width Approximation ($\eta=1$) and  is discussed in detail in Appendix \ref{app2}. Here $X= LV,\ LC$ and $\Gamma_{\rm Ma}(M_{N})=(\Gamma_{\rm Ma}(M_{N_1})+\Gamma_{\rm Ma}(M_{N_2}))/2$ is the average heavy-neutrino total-decay-width.  The kinematical functions $\widetilde{\Gamma}_{LV}$ and $\widetilde{\Gamma}_{LC}$ can be written as
\begin{align}
\label{fact}
\widetilde{\Gamma}_{LV}&(M_N)  \equiv \widetilde{\Gamma}_{LC}(M_N)   \equiv  \widetilde{\Gamma}\big( W^+ \to \ell^+_1 N \big) \  \widetilde{\Gamma}\big( N \to \ell^+_2 {\ell}^{' -} \bar{\nu}_{\ell^{'}}  \big) \ ,
\end{align}
where the quantities with tilde $\widetilde{\Gamma}$ do not have explicit dependence on the mixing elements $|B_{\ell N_i}|$ and are given by
\begin{align}
 \label{W2bardecay}
 \widetilde{\Gamma}\big( W^+ \to \ell^+_1 N \big) &= \frac{G_F}{12 \pi \sqrt{2} M_W} \lambda^{\frac{1}{2}}\Bigg(1,\frac{M_N^2}{M_{W}^2},\frac{m_1^2}{M_{W}^2}\Bigg) \\
 \nonumber &\times \Big( 2 M_W^4-M_W^2 M_N^2-M_N^4  + m_1^2 \big(-M_W^2+2 M_N^2-{m_1}^2 \big) \Big) \ , \\
\nonumber \\ 
 \label{N3bardecay}
   \widetilde{\Gamma} \big( N \to \ell^+_2 \ell^{' -} \bar{\nu_{\ell^{'}}}  \big) &= \frac{ G_F^2 M_W^5}{2 (4 \pi)^3} \; {\overline {\rm Int}} \ .
\end{align}
The dimensionless quantity ${\overline {\rm Int}}$ introduced in Eq.~\eqref{N3bardecay} is
\bea
 {\overline {\rm Int}} & =& \frac{1}{M_W} \int_{{m_{\ell^{'}}}}^{(E_{\ell^{'}})_{\rm max}} d E_{\ell^{'}} {\bigg \{}
{\bar A}_2 \left[ z_{\rm max}(E_{\ell^{'}}) - z_{\rm min}(E_{\ell^{'}}) \right] + {\bar A}_1 \frac{1}{2} \ln \left[ \frac{1 + z_{\rm max}(E_{\ell^{'}})^2 }{1 + z_{\rm min}(E_{\ell^{'}})^2 } \right] 
\nonumber\\ &&
+ {\bar A}_0(E_{\ell^{'}}) \left[ {\rm ArcTan}( z_{\rm max}(E_{\ell^{'}}) ) - {\rm ArcTan}( z_{\rm min}(E_{\ell^{'}}) ) \right] {\bigg \}},
\label{barInt}
\eea
where $(E_{\ell^{'}})_{\rm max}$, is the maximal value of $E_{\ell^{'}}$ in the rest system of heavy neutrino
\be
(E_{\ell^{'}})_{\rm max} = \frac{1}{2 M_N} (M_N^2 - {m_2}^2 + {m_{\ell^{'}}}^2),
\label{Elmax}
\ee
and the dimensionless coefficients $z_{\rm max}(E_{\ell^{'}})$, $z_{\rm min}(E_{\ell^{'}})$ and ${\bar A}_j$ ($j=0,1,2$) in the integrand of Eq.~(\ref{barInt}) are given in App.~\ref{app3}. We noticed that when $M_N \ll M_W$ and $m_{\ell^{'}}=0$ the Eq.~\eqref{N3bardecay} can be approximated as
\begin{align}
\nonumber   \widetilde{\Gamma} \big( N & \to \ell^+_2 {\ell}^{' -} \bar{\nu}_{\ell^{'}}  \big) = \frac{ G_F^2}{192 \pi^3 M_N^3}  \\
   &\times  \Bigg[ M_N^8-m_2^8 + 8 m_2^2 M_N^2 \big( m_2^4-M_N^4 \big) + 24 m_2^4 M_N^4 \log \Bigg( \frac{M_N}{m_2} \Bigg) \Bigg] \ ,
\end{align}
and when, in addition, $m_2=0$, this expression simplifies further
\begin{equation}
{\widetilde \Gamma}(N \to \ell_2^{+} \ell^{' -} {\bar \nu}_{\ell^{'}}) =
\frac{G_F^2}{192 \pi^3} M_N^5.
\end{equation}

\begin{figure}[H]
\centering
\includegraphics[scale = 0.77]{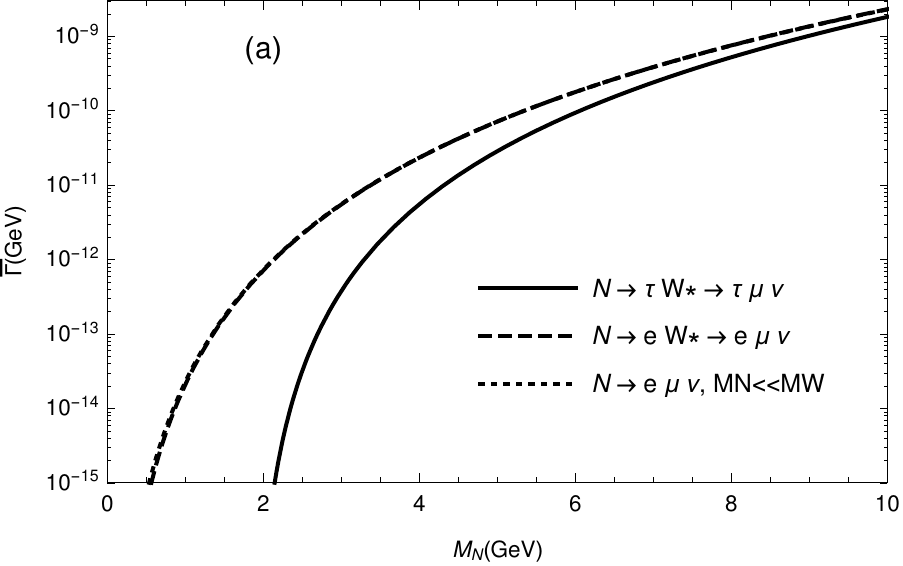}\hspace{0.5 cm}
\includegraphics[scale = 0.77]{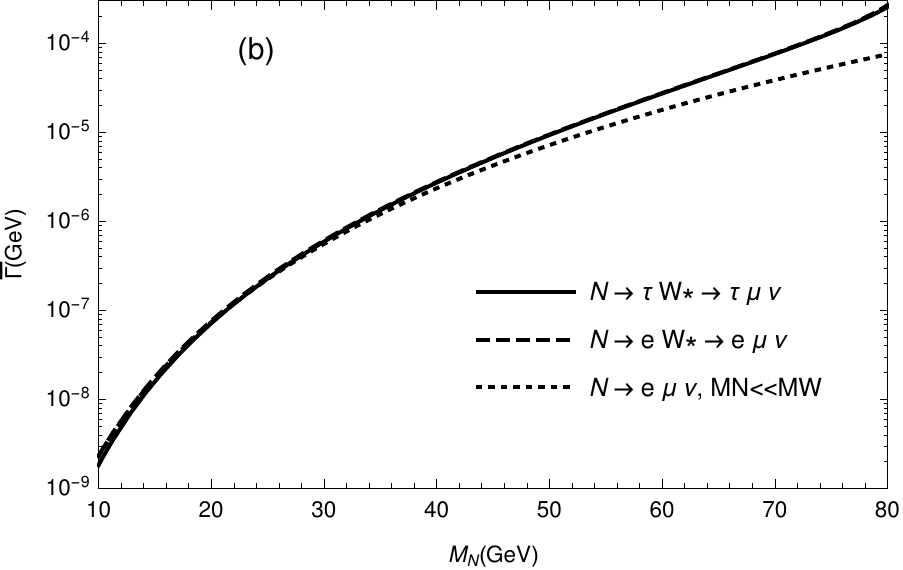}
\caption{Decay width ${\widetilde \Gamma}(N \to \ell_2^+ \ell^- {\bar \nu}_{\ell})$ as a function of $M_N$, for two different charged leptons in the final state ($\ell_2 = \tau, \mu$). The panel (a) stands for $M_N \leq 10$ and the (b) for $10\leq M_N \leq 80$.}
\label{fig:GammaBarNl2lnu}
\end{figure}

In Figs. \ref{fig:GammaBarNl2lnu}(a)-\ref{fig:GammaBarNl2lnu}(b) we present the decay width ${\widetilde \Gamma}(N \to \ell_2^+ \ell^{' -} {\bar \nu}_{\ell^{'}})$ as a function of $M_N$, for two different choices $\ell_2 = \tau, \mu$, and compare it with the $M_N \ll M_W$ approximation. We note that the latter approximation is widely used in the literature, i.e., the $W$-propagator in the decay is usually taken as a constant $\propto 1/M_W^2$. We see from Fig. \ref{fig:GammaBarNl2lnu}(b) that the latter approximation is clearly not justified when $M_N > 40$ GeV

By replacing Eq.~\eqref{W2bardecay} and Eq.~\eqref{N3bardecay} in Eq.~\eqref{fact} we find
\begin{align}
\nonumber
\widetilde{\Gamma}\big( W^+ \to \ell^+_1 N \big)& \;  \widetilde{\Gamma}\big( N \to \ell^+_2 {\ell}^{' -} \bar{\nu}_{\ell^{'}}  \big) =  \frac{G_F^3 M_W^4}{1536 \pi^4 \sqrt{2}}\; \lambda^{\frac{1}{2}}\Bigg(1,\frac{M_N^2}{M_{W}^2},\frac{m_1^2}{M_{W}^2}\Bigg) \times {\overline {\rm Int}}  \\
& \times \Big( 2 M_W^4-M_W^2 M_N^2-M_N^4  + m_1^2 \big(-M_W^2+2 M_N^2-{m_1}^2 \big) \Big)   \ .
\end{align}
It is important to remark that differences between $\Gamma_{LV}(W^{\pm})$ and $\Gamma_{LC}(W^{\pm})$  manifests themselves only in the {\textit CP phases} ($\theta_{LV}$ and $\theta_{LC}$) present in Eq.~\eqref{DWE}.  Finally $\lambda^{\frac{1}{2}}(x,y,z)$, is
\begin{equation}
\lambda^{\frac{1}{2}}(x,y,z)=\Big( x^2 + y^2 + z^2 -2 xy -2 xz -2 yz \Big)^{\frac{1}{2}} .
\end{equation}

\subsection{The effects of heavy neutrino oscillations}

\begin{figure}[H]
\centering
\includegraphics[scale = 0.47]{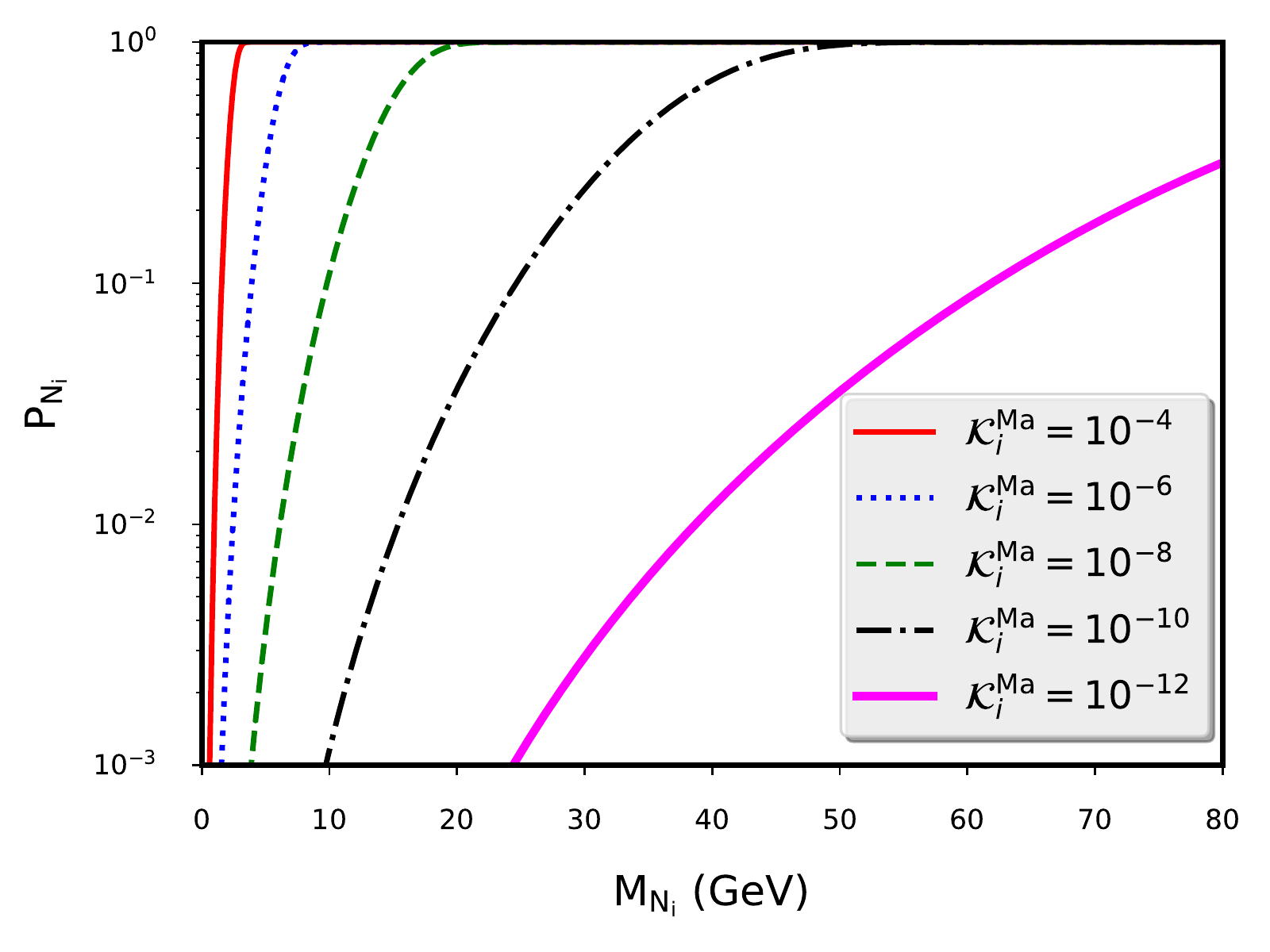}
\includegraphics[scale = 0.47]{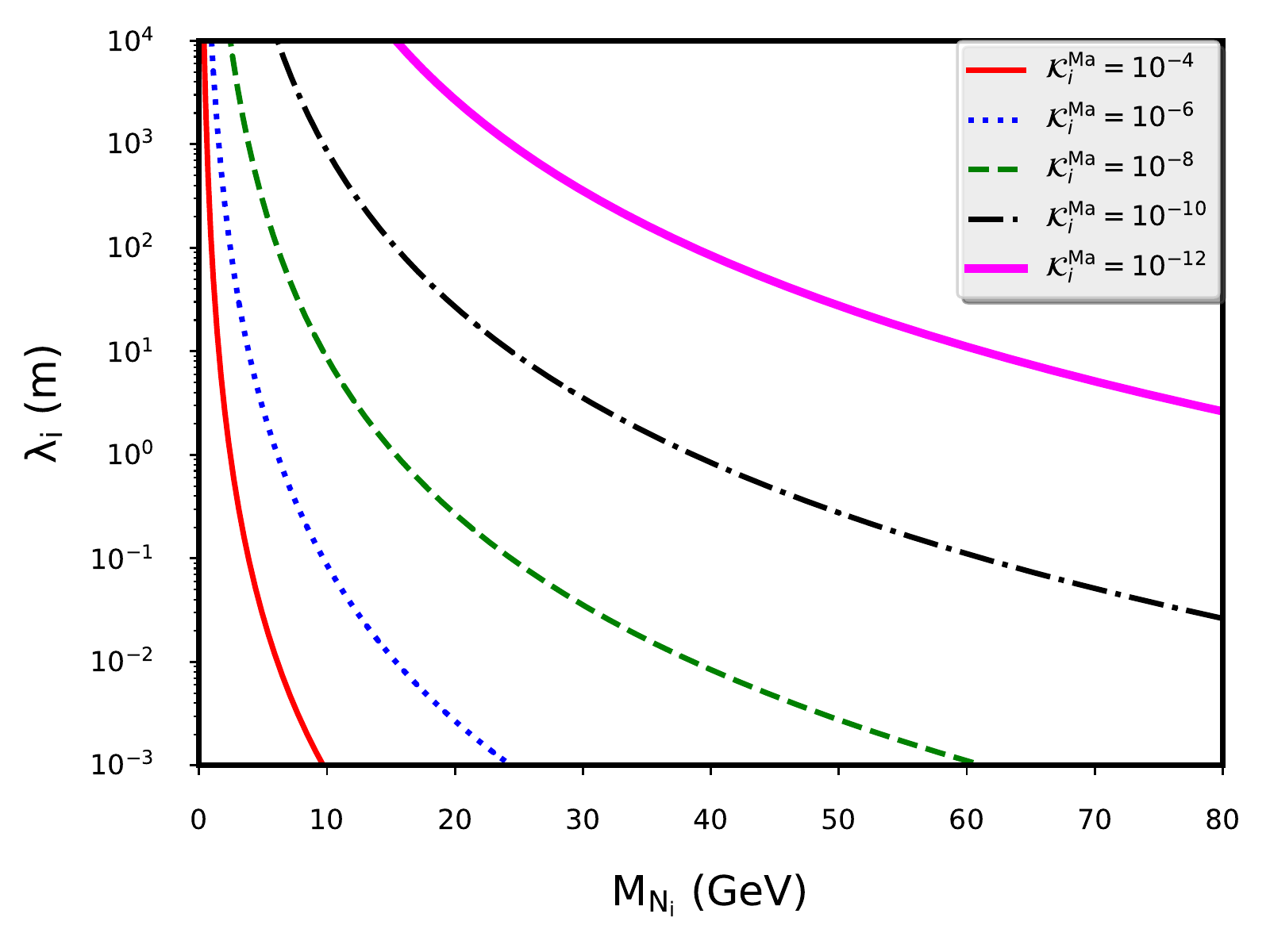}
\caption{Acceptance factor $P_{N_i}$ as a function of the mass (left panel) and decay length $\lambda_i=\beta_{N_i}\gamma_{N_i} / \Gamma_{N_i}$ as a function of the mass (right panel). Here we have used $\gamma_{N_i} \beta_{N_i}=2$.}
\label{fig:PN}
\end{figure}

In order to account for the effects of heavy neutrino oscillations, we will follow the approach for two flavor presented in Ref.~\cite{Cvetic:2015ura}. From now on, we will consider the case when $\ell_1 = \mu$, $\ell_2 = \tau$ and $\ell^{'} = e$ ($m_e \approx 0$), the heavy neutrinos $N_1$ and $N_2$ are almost degenerate and simultaneously the degeneracy level ($|\Delta M_N|$) is significantly larger than the neutrino decay width $\Gamma_{N}$, which is very small.  In addition, for the oscillation to be well defined, we require that the heavy neutrino decay after one or more oscillation cycles. It means that decay length $\lambda_i= \beta_{N_i} \gamma_{N_i} / \Gamma_{N_i}$, which is the typical  distance between the
vertex where N is produced and the vertex where it decays, must be much greater than oscillation length $L_{\rm osc} = \frac{2 \pi \beta_N \gamma_N}{\Delta M_N}$. At this point we mention that heavy neutrino-antineutrino oscillation has been studied for the low scale seesaw model in \cite{Antusch:2016ejd} considering long lived heavy neutrinos so that displaced decay of the heavy neutrino can be observed. The aforementioned sentence can be summarized as
\begin{equation}
\label{cond}
|\Delta M_N| \ll M_{N_1} \quad  \quad {\rm and} \quad \quad  y \equiv \frac{|\Delta M_N|}{\Gamma_{N}} \gg 1 \ .
\end{equation}
It is clear from Fig.~\ref{delta_eta} that under the conditions presented in Eq.~(\ref{cond})  the factors $\delta_{12}$ and $\frac{\eta(y)}{y}$ becomes very small and can be neglected. Furthermore, the heavy neutrino masses can be regarded as $M_N \equiv M_{N_1} \approx M_{N_2}$. As a result, the decay width Eq.~\eqref{DWE} is reduced to the following form
\begin{align}
\label{DWEred}
\nonumber \Gamma_{LV}(W^{\pm})& \equiv \Gamma_{LC}(W^{\pm}) = \\
& =\widetilde{\Gamma}\big( W^+ \to \ell^+_1 N \big) \ \widetilde{\Gamma}\big( N \to \ell^+_2 e^- \bar{\nu}  \big) \Bigg(  \frac{|B_{\mu N_1}^{}|^2 |B_{\tau N_1}^{}|^2}{\Gamma_{\rm Ma}(M_{N_1})}+ \frac{|B_{\mu N_2}^{}|^2 |B_{\tau N_2}^{}|^2}{\Gamma_{\rm Ma}(M_{N_2})} \Bigg) \ .
\end{align}
It is important to point out that Eq.~(\ref{DWEred}) does not yet contain the acceptance factor $P_N$, which accounts for the fact that the decay will be detected only if the on-shell neutrino decays during his passage through the detector. The probability $P_{N_i}$ as function of the distance $L$ of decay of $N_i$ is
\begin{align}
P_{N_i}(L) = 1-\exp\Bigg[-\frac{L \ \Gamma_{\rm Ma}(M_{N_i})}{\gamma_{N_i} \ \beta_{N_i}}\Bigg] \approx \frac{L \Gamma_{N}(M_{N_i})}{\gamma_{N_i} \beta_{N_i}} \ .
\label{PN}
\end{align}
This approximation is valid when $\Gamma_{\rm Ma}(M_{N_i}) \ll 1$. In Eq. \eqref{PN} $\gamma_{N_i} = E/M_{N_i} \equiv \gamma_N$ is the relativistic Lorentz factor, and $\beta_{N_i} = v_{N_i}/c \equiv \beta_N$. 
Although the values of $\gamma_N \beta_N$ of the considered neutrinos $N_j$ at LHC can be as high as $\sim 10^2$, only the values of $\gamma \beta \sim 10^0$ are relevant for our considerations, because the differential decay width $d \Gamma^{\rm osc}(W^{\pm})/dL$ are proportional to $1/(\gamma_N \beta_N)$, as seen in Eq.~\eqref{effdifdw}.
As we can see from Eq.~(\ref{DNwidth}), the factor $\Gamma_{\rm Ma}(M_{N_i})$ depends on the coefficient $\K^{\rm Ma}_i$ which is ${\cal K}_i^{\rm Ma} \stackrel{<}{\sim} 2 \times 10^{-4}$ according to the present mixing limits\footnote{The present mixing limits are presented in Fig.~\ref{fig:mix1}  and Fig.~\ref{fig:mix2} and the corresponding effective mixing coefficients in Fig.~\ref{fig:efcoef}.}. In Fig.~\ref{fig:PN} we present the values for $P_{N_i}$ and decay length $\lambda_i$ for different $\K^{\rm Ma}_i$. Besides these, in Table.~\ref{valPNapp} we present the range of applicability of the approximation shown on the right-hand side of Eq.~\eqref{PN}
\begin{table}[t]
\centering 
\begin{tabular}{|c | c | c | c | c | c |} 
\hline\hline 
{\footnotesize $\mathcal{K}^{\rm Ma}$} & {\footnotesize $2 \leq M_N \leq 4  $ } & {\footnotesize $ 4 \leq M_N \leq 10  $ } & {\footnotesize $10 \leq M_N \leq 25 $ } & {\footnotesize $ 25 \leq M_N \leq 80$  } \\ [0.5ex] 
\hline 
$10^{-4}$ & No & No & No  & No \\ 
$10^{-6}$ & Yes & No & No  & No \\
$10^{-8}$ & Yes & Yes & No  & No \\
$10^{-10}$ & Yes & Yes & Yes  & No \\
$10^{-12}$ & Yes & Yes & Yes & Yes \\ [1ex] 
\hline 
\end{tabular}
\caption{Validity of approximation shown on the right-hand side of Eq.~\eqref{PN} for $\mathcal{K}^{\rm Ma}_i$ presented in Fig.~\ref{fig:PN}. Here {\bf Yes} stand when the approximation is applicable and {\bf No} when is not.} 
\label{valPNapp} 
\end{table}
Combining the acceptance factor Eq.~\eqref{PN} and the decay width shown in Eq.~\eqref{DWEred} we get the effective decay width
\begin{align}
\label{DWeff}
\nonumber \Gamma_{LV}^{\rm eff}(W^{\pm})& \equiv \Gamma_{LC}^{\rm eff}(W^{\pm}) = \widetilde{\Gamma}\big( W^+ \to \ell^+_1 N \big) \ \widetilde{\Gamma}\big( N \to \ell^+_2 e^- \bar{\nu}  \big)  \times \\
&\Bigg( 1-\exp\Big[-\frac{L \ \Gamma_{\rm Ma}(M_{N})}{\gamma_N \ \beta_N}\Big]  \Bigg)  \Bigg(  \frac{|B_{\mu N_1}^{}|^2 |B_{\tau N_1}^{}|^2}{\Gamma_{\rm Ma}(M_{N_1})}+ \frac{|B_{\mu N_2}^{}|^2 |B_{\tau N_2}^{}|^2}{\Gamma_{\rm Ma}(M_{N_2})} \Bigg) \ .
\end{align}
We noticed that, even if the coefficients $\mathcal{N}_{\ell i}^{\rm Ma}$ in $\mathcal{K}^{\rm Ma}_i$ are common for both neutrinos, the factor $P_{N_i}$, is not the same because the mixings $B_{\ell N_1}$ and $B_{\ell N_2}$ can be, in principle, different for the two neutrinos, and therefore the two mixing factors $\mathcal{K}^{\rm Ma}_i$ $(i = 1, 2)$ may differ significantly from each other. However, from now on we will assume that  $\mathcal{K}^{\rm Ma}_1 \approx \mathcal{K}^{\rm Ma}_2$. Hence, the effective differential decay widths, with respect to the distance L between the two vertices of the processes, are
\begin{align}
\label{effdifdw}
\nonumber \frac{d}{dL}\; \Gamma_{X}^{\rm eff}(W^{\pm})  & =  \frac{1}{\gamma_N \ \beta_N}\; \exp\Big[-\frac{L \ \Gamma_{\rm Ma}(M_N)}{\gamma_N \ \beta_N}\Big] \;  \widetilde{\Gamma}\big( W^+ \to \ell^+_1 N \big) \ \widetilde{\Gamma}\big( N \to \ell^+_2 e^- \bar{\nu}  \big) \\
& \quad \quad \times  \Bigg(|B_{\mu N_1}^{}|^2 |B_{\tau N_1}^{}|^2+ |B_{\mu N_2}^{}|^2 |B_{\tau N_2}^{}|^2 \Bigg)\ .
\end{align}
Up to now, we have not considered the $N_1 - N_2$ oscillations effects. As  we mentioned before, these effects have been studied in detail in Ref.~\cite{Cvetic:2015ura} and as there, it is straightforward to show here that, for the $W_X^{\pm}$ decay, the L dependent effective differential decay width considering heavy neutrinos is
\begin{small}
\begin{align}
\nonumber \frac{d}{dL}\;& \Gamma_{X}^{\rm osc}(W^{\pm})  = \frac{ 1}{\gamma_N \ \beta_N} \exp\Big[-\frac{L \ \Gamma_{\rm Ma}(M_N)}{\gamma_N \ \beta_N}\Big] \; \widetilde{\Gamma}\big( W^+ \to \ell^+_1 N \big) \ \widetilde{\Gamma}\big( N \to \ell^+_2 e^- \bar{\nu}  \big)  \\
& \times \Bigg( \sum_{i=1}^2 |B_{\mu N_i}|^2 |B_{\tau N_i}|^2+ 2 |B_{\mu N_1}| |B_{\tau N_1}| |B_{\mu N_2}| |B_{\tau N_2}| \cos \Big(2\pi \; \frac{L}{L_{\rm osc}} \pm \theta_{X} \Big)\Bigg) \ , 
\label{effdwfosc}
\end{align}
\end{small}
where $L_{\rm osc} = \frac{2 \pi \beta_N \gamma_N}{\Delta M_N}$ and $X=LV, \ LC$. Integrating the Eq.~\eqref{effdwfosc} over $dL$ up to the full length\footnote{The full length $L$, in principle, could be the detector length.} $L$, we find
\begin{small}
\begin{align}
\label{oscWfulldw}
\nonumber \Gamma_{X}^{{\rm osc}}(W^{\pm}) =  & \frac{\widetilde{\Gamma}\big( W^+ \to \mu^+ N \big) \ \widetilde{\Gamma}\big( N \to \tau^+ e^- \bar{\nu}_e  \big)}{\Gamma_{\rm Ma}(M_{N})}  {\Bigg [}
\left( 
1 - \exp \left[ - \frac{L \Gamma_{\rm Ma}(M_N)}{\gamma_N \beta_N} \right] 
\right)
\times
\nonumber\\ &
\sum_{i=1}^{2} |B_{\mu N_i}|^2 |B_{\tau N_i}|^2
+ \frac{2}{y^2} |B_{\mu N_1}| |B_{\tau N_1}|  |B_{\mu N_2}| |B_{\tau N_2}|
\nonumber\\ &
{\Big (}  
\cos (\theta_X)  \mp  y \sin (\theta_X) 
+ \exp \left[ - \frac{L \Gamma_{\rm Ma}(M_N)}{\gamma_N \beta_N} \right] \times
\nonumber\\ &
\left[
y \sin \left( 2 \pi \frac{L}{L_{\rm osc}} \pm \theta_X \right) -
   \cos \left( 2 \pi \frac{L}{L_{\rm osc}} \pm \theta_X \right) 
\right]
{\Big )}
{\Bigg ]}.
\end{align}
\end{small}
In Eq.~(\ref{oscWfulldw}) we neglected the relative corrections ${\cal O}( (L_{\rm osc}/{\bar L})^2 ) = {\cal O}(1/y^2)$. Assuming $|B_{\mu N_i}|^2 \approx |B_{\tau N_i}|^2 \equiv |B_{\ell N}|^2$ ($i=1,2$) and $\mathcal{K}^{\rm Ma}_1 \approx \mathcal{K}^{\rm Ma}_2 \approx 10 |B_{\ell N}|^2$, Eq.~\eqref{oscWfulldw} becomes 
\begin{small}
\begin{align}
\label{oscWfulldwapp}
\nonumber \Gamma_{X}^{{\rm osc}}(W^{\pm})& = |B_{\ell N}|^2  F\big[W\big]   \Bigg[ \Bigg(1-\exp\Big[-\frac{L \ \Gamma_{\rm Ma}(M_N)}{\gamma_N \ \beta_N}\Big] \Bigg) +   \\
\nonumber & \quad \quad   \frac{1}{y^2} \; \times \Bigg(  cos(\theta_X) \mp  y \sin(\theta_X) + \exp\Big[-\frac{L \ \Gamma_{\rm Ma}(M_N)}{\gamma_N \ \beta_N}\Big] \times  \\
&  \quad \quad \Bigg[ y \sin\Big(2\pi \; \frac{L}{L_{\rm osc}} \pm \theta_{X} \Big)  
-\cos \Big(2\pi \; \frac{L}{L_{\rm osc}} \pm \theta_{X} \Big) \Bigg]  \Bigg) 
\Bigg] \ ;
\end{align}
\end{small}
where
\begin{align}
\nonumber
F\big[W\big] &=  \frac{G_F M_W^4}{80 \pi \sqrt{2} M_N^5}\; \lambda^{\frac{1}{2}}\Bigg(1,\frac{M_N^2}{M_{W}^2},\frac{m_1^2}{M_{W}^2}\Bigg) \times {\overline {\rm Int}}  \\
& \times \Big( 2 M_W^4-M_W^2 M_N^2-M_N^4  + m_1^2 \big(-M_W^2+2 M_N^2-{m_1}^2 \big) \Big)   \ .
\end{align}

Finally, the branching ratio including the heavy neutrino oscillations effects are
\begin{align}
\label{oscWfullBR}
{\rm Br}_{X}^{{\rm osc}}(W^{\pm})& = \frac{\Gamma_{X}^{{\rm osc}}(W^{\pm})}{\Gamma(W^{+})\to {\rm all}}  \ .
\end{align}

The above analysis can be extended to the top quark decays (see App.~\ref{app4} for details).
\section{Discussion and Results}
 
In this Section we use the main results obtained in Sec.~\ref{s3} in order to provide suggestions for future searches. The upcoming experiments at the High-Luminosity Large Hadron Collider (HL-LHC), proposed at CERN \cite{Apollinari:2017cqg}, are expected to collect $\sim 10^{11}$ $W^{\pm}$ bosons \cite{Mangano:2014xta}. Such a huge amount of $W^{\pm}$ particles opens a new window to explore the rare $W^{\pm}$ decay parameters.

\begin{figure}[H]
\centering
\includegraphics[scale = 0.45]{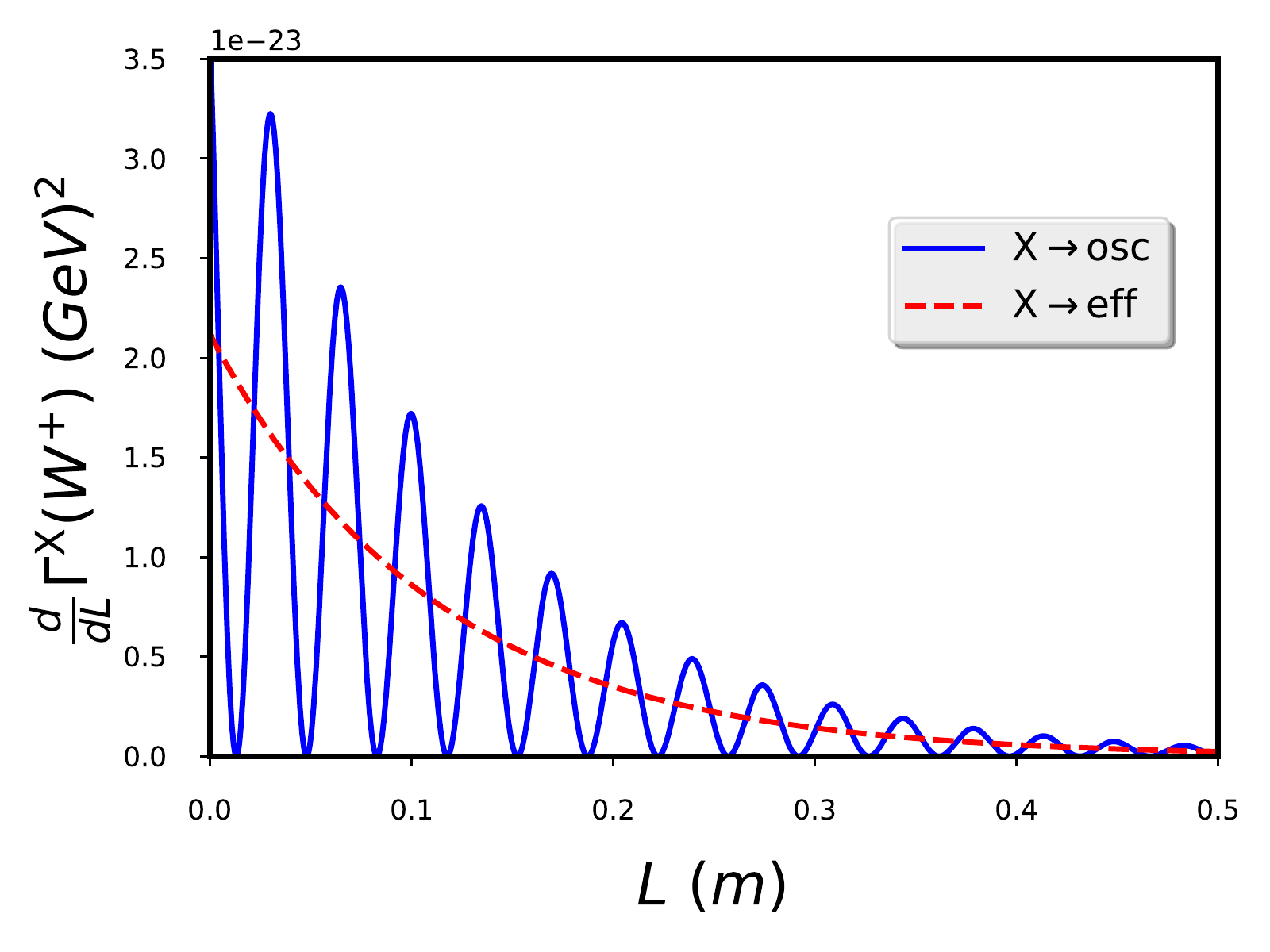}\hspace{0.3 cm}
\includegraphics[scale = 0.45]{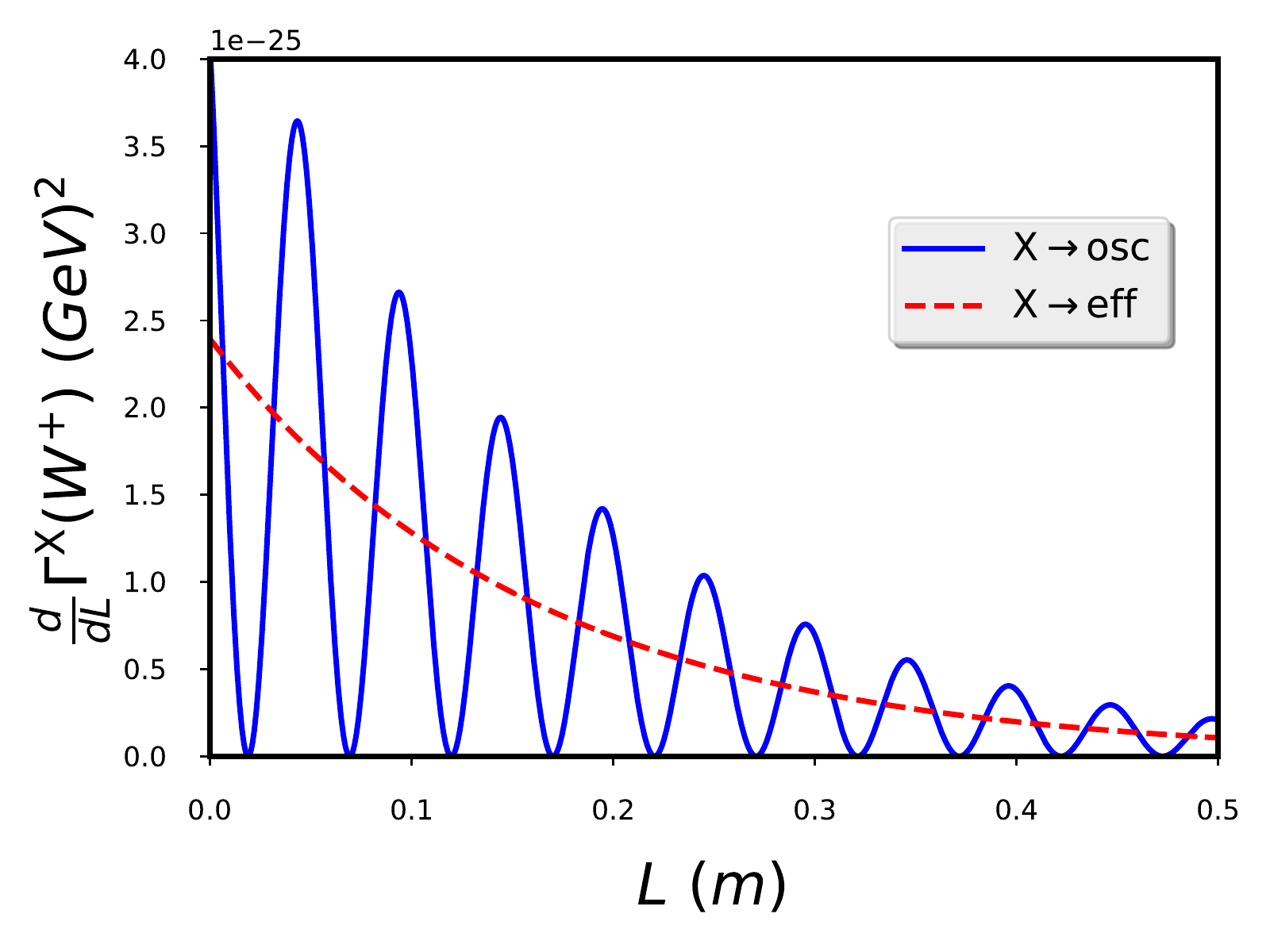}
\caption{Differential decay width $\frac{d  \Gamma^{\rm osc}(W^{+})}{dL}$ as a function of $L$ for different mixing and masses. Left panel: $|B_{\ell N}|^2 \approx \mathcal{K}^{\rm Ma}/10=10^{-6}$ and $M_N=6$ GeV. Right panel: $|B_{\ell N}|^2 \approx \mathcal{K}^{\rm Ma}/10=10^{-8}$ and $M_N=14$ GeV. Here we used $\gamma_N \beta_N=2$ and $y=20$.}
\label{fig:diffGammaL}
\end{figure}

In Fig.~\ref{fig:diffGammaL} we present, for a representative choice of parameters, comparison between the differential decay width expected for $W^+$ decays as a function of the detector length $L$: neglecting the oscillation effects in Eq.~\eqref{effdifdw} and considering them in Eq.~\eqref{effdwfosc}.

The presence of oscillating terms in Eq.~\eqref{effdwfosc} is manifested in the decay width Eq.~\eqref{oscWfulldw} as a modulation. The existence of such oscillating terms, if detected separately for $W^-$ and $W^+$, may even allow us to explore, as a secondary result, the CP-phases ($\theta_{LV}$ and $\theta_{LC}$) for a given neutrino mass $M_N$ and mixings $|B_{\ell N}|^2$, cf.~(Eq.~\eqref{effdwfosc}). We recall that the factor $\sin \theta_X$ appears in the CP asymmetry factor ${\cal A}_{\rm CP}$
\begin{eqnarray}
{\cal A}_{\rm CP}^{(X)}(W) &\equiv& \frac{ \Gamma_X(W^-) -  \Gamma_X(W^+)}{ \Gamma_X(W^-) +  \Gamma_X(W^+)}
\nonumber\\
 &\propto& P \frac{y}{(y^2+1)} \sin \theta_X ,
\label{ACP}
\end{eqnarray}
where the proportionality factor $P \sim 1$ depends on ratios of mixings.

\begin{figure}[H]
\centering
\includegraphics[scale = 0.45]{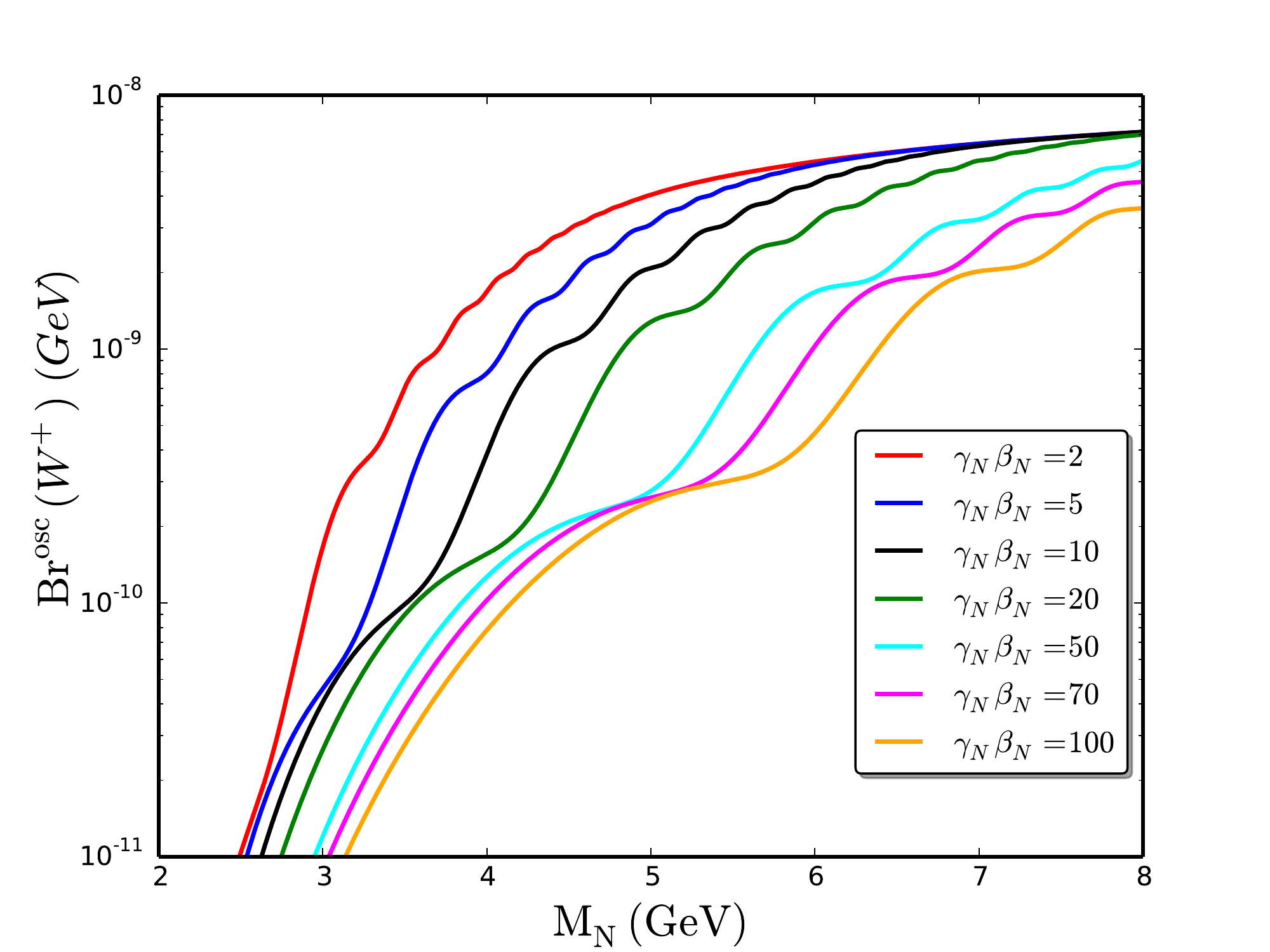}
\caption{Branching ratios ${\rm Br}^{{\rm osc}}(W^{+})$ for  $L=1\ m$, $\theta = \pi/3 $, $|B_{\ell N}|^2 \approx \mathcal{K}^{\rm Ma}/10=10^{-6}$ and $y=20$  as a function of $M_N$ for different $\gamma_N \beta_N$.  The higher curves correspond to lower values of $\gamma_N \beta_N$.}
\label{fig:diffGamma}
\end{figure}

\begin{figure}[H]
\centering
\includegraphics[scale = 0.28]{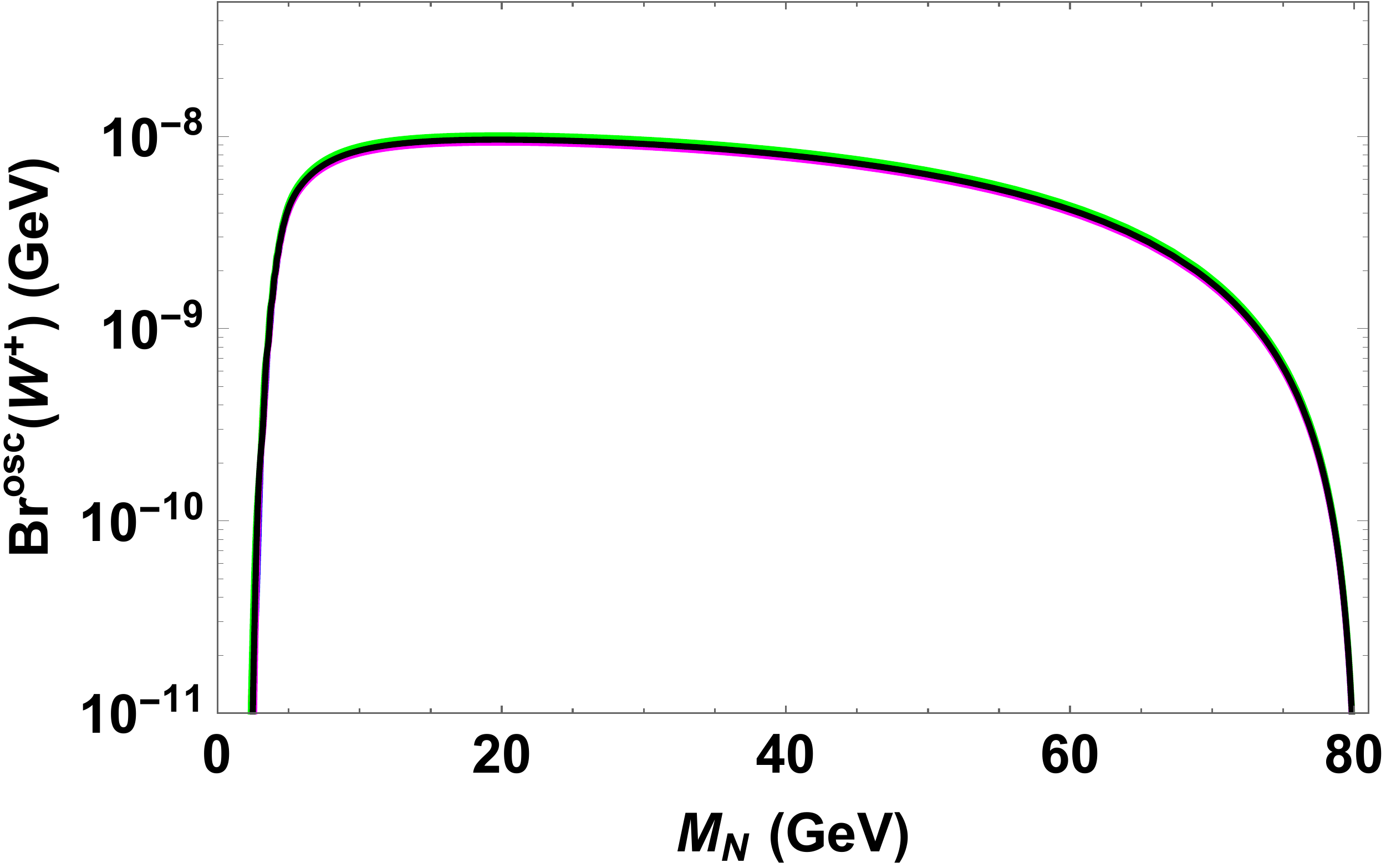}\hspace{0.3 cm}
\includegraphics[scale = 0.28]{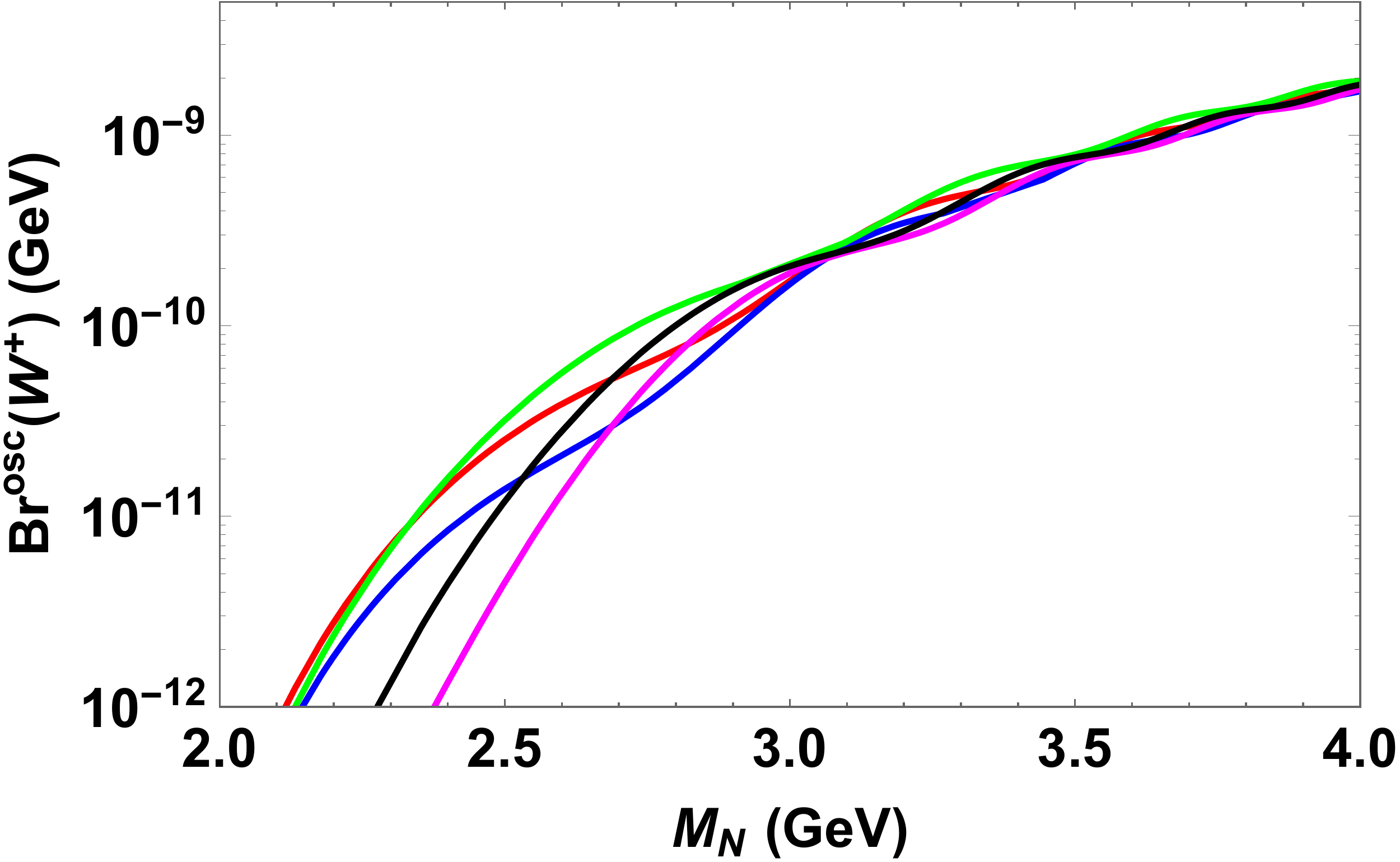} \\ \vspace{0.5 cm}
\includegraphics[scale = 0.28]{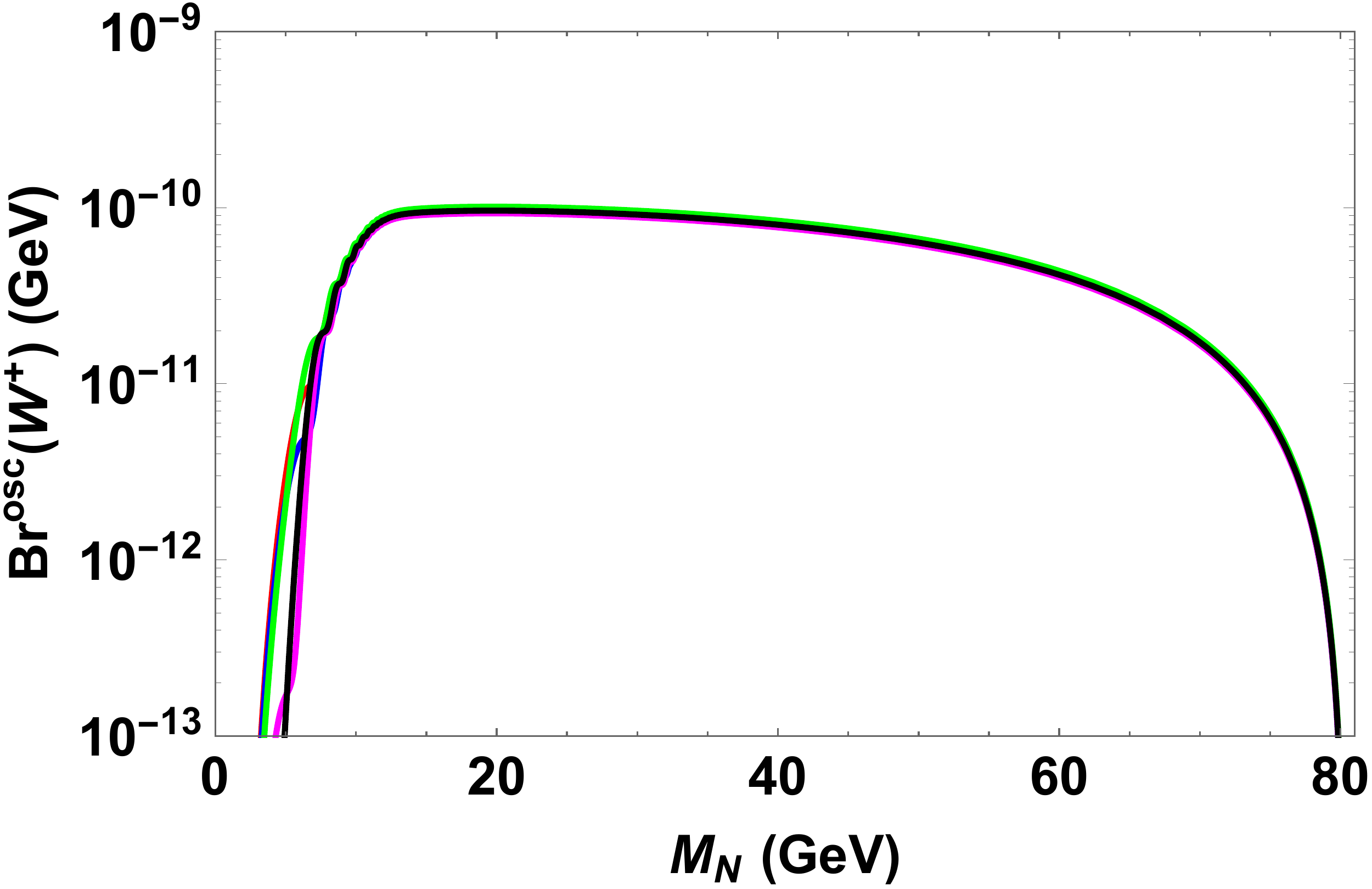}\hspace{0.3 cm}
\includegraphics[scale = 0.28]{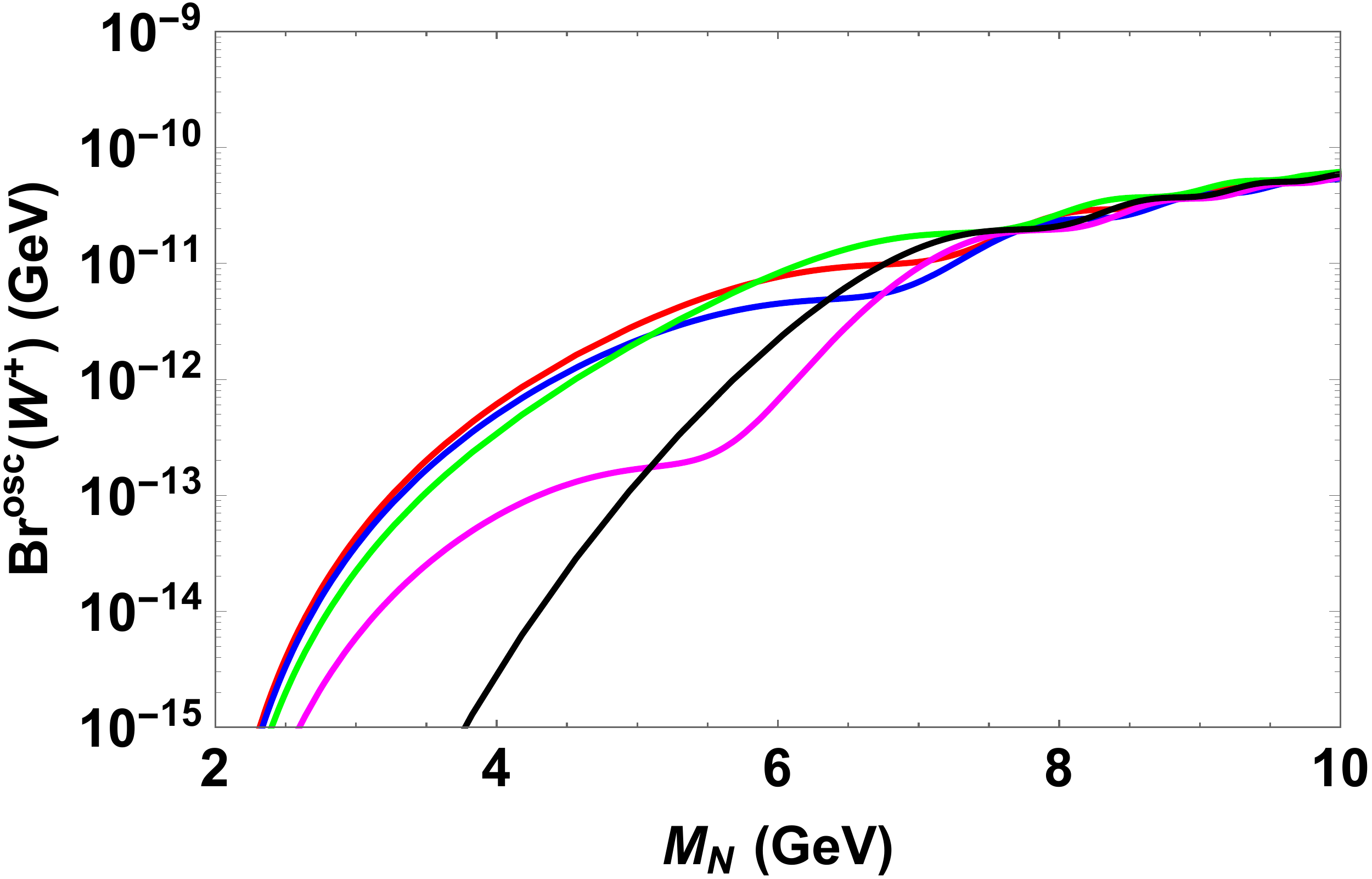}
\caption{Branching ratios ${\rm Br}^{{\rm osc}}(W^{+})$ for $\gamma_N \beta_N=2$,  $L=1\ (m)$ and $y=20$  as a function of $M_N$ for different $\theta$ angles. The red line stands for $\theta = 0$, blue for $\theta = \pi/4$, green for $\theta = \pi/2$, magenta for $\theta = 3 \pi /4$ and the black for $\theta = \pi$. Top: we provided  $|B_{\ell N}|^2 \approx \mathcal{K}^{\rm Ma}/10=10^{-6}$. Bottom: we used $|B_{\ell N}|^2 \approx \mathcal{K}^{\rm Ma}/10=10^{-8}$.}
\label{fig:BroscWpMN}
\end{figure}

In Figs.~\ref{fig:diffGamma} and ~\ref{fig:BroscWpMN} we present the branching ratios ($\propto$ decay widths) as a function of $M_N$, for a given distance between the vertices $L =1 \ m$ and for $y$ ($\equiv |\Delta M|/\Gamma_N$)$=20$: in Fig.~\ref{fig:diffGamma} for fixed $|B_{\ell N}|^2=10^{-6}$, $\theta=\pi/3$ and various fixed values of $\gamma_N \beta_N$; in Figs.~\ref{fig:BroscWpMN} for $\gamma_N \beta_N=2$, $|B_{\ell N}|^2=10^{-6}$ and $10^{-8}$ and various fixed values of $\theta$.  Keeping in mind that $\sim 10^{11}$ $W^{\pm}$ will be collected at HL-LHC, Figs.~\ref{fig:BroscWpMN} suggest that the modulation of ${\rm Br}^{\rm osc}(W^+)$  is appreciable and possibly detectable only when $2.5 \ {\rm GeV} < M_N < 3.5$ GeV if $|B_{\ell N}|^2=10^{-6}$, and the rates are in general too low if $|B_{\ell N}|^2=10^{-8}$

\begin{figure}[H]
\centering
\includegraphics[scale = 0.28]{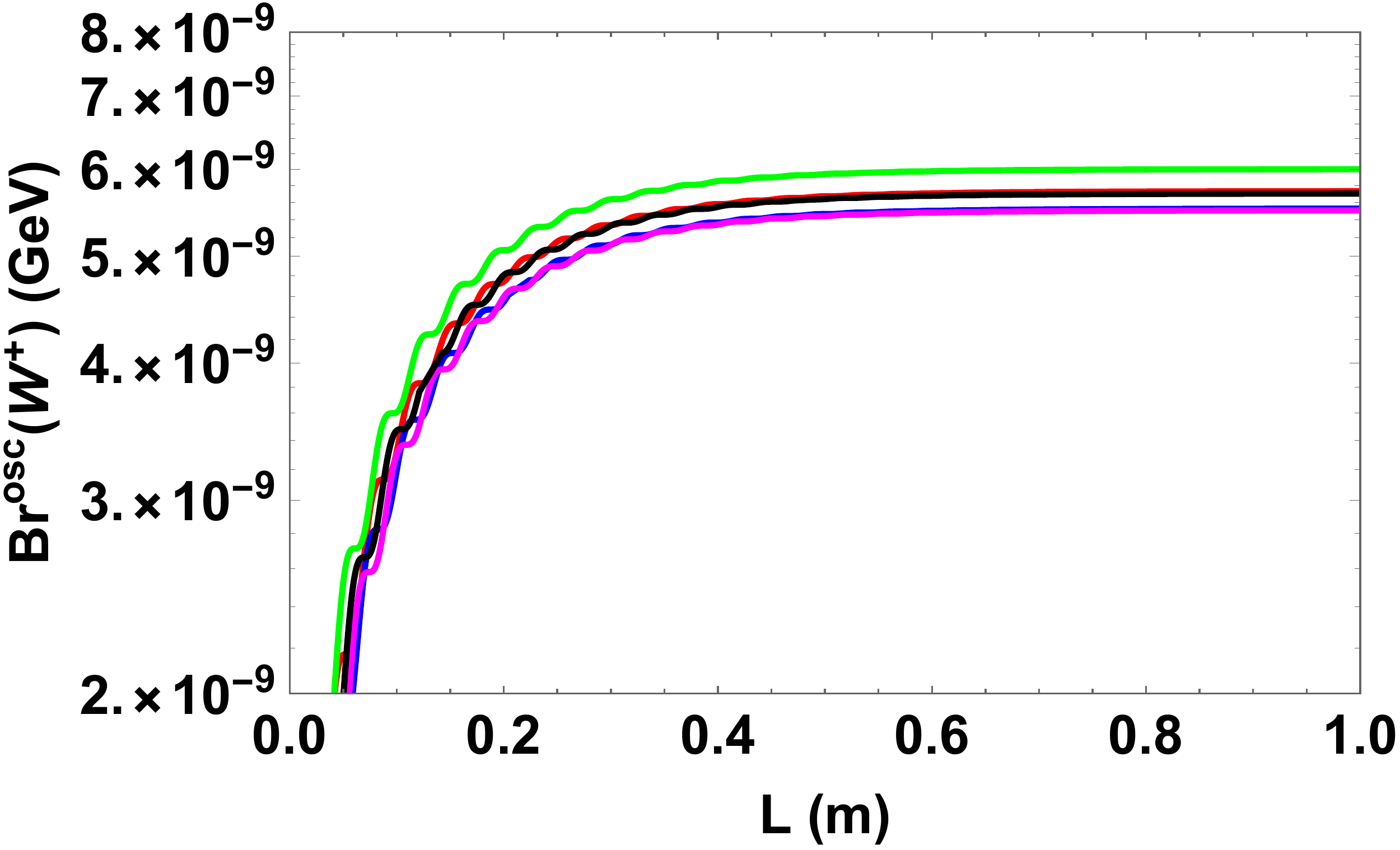}\hspace{0.3 cm}
\includegraphics[scale = 0.28]{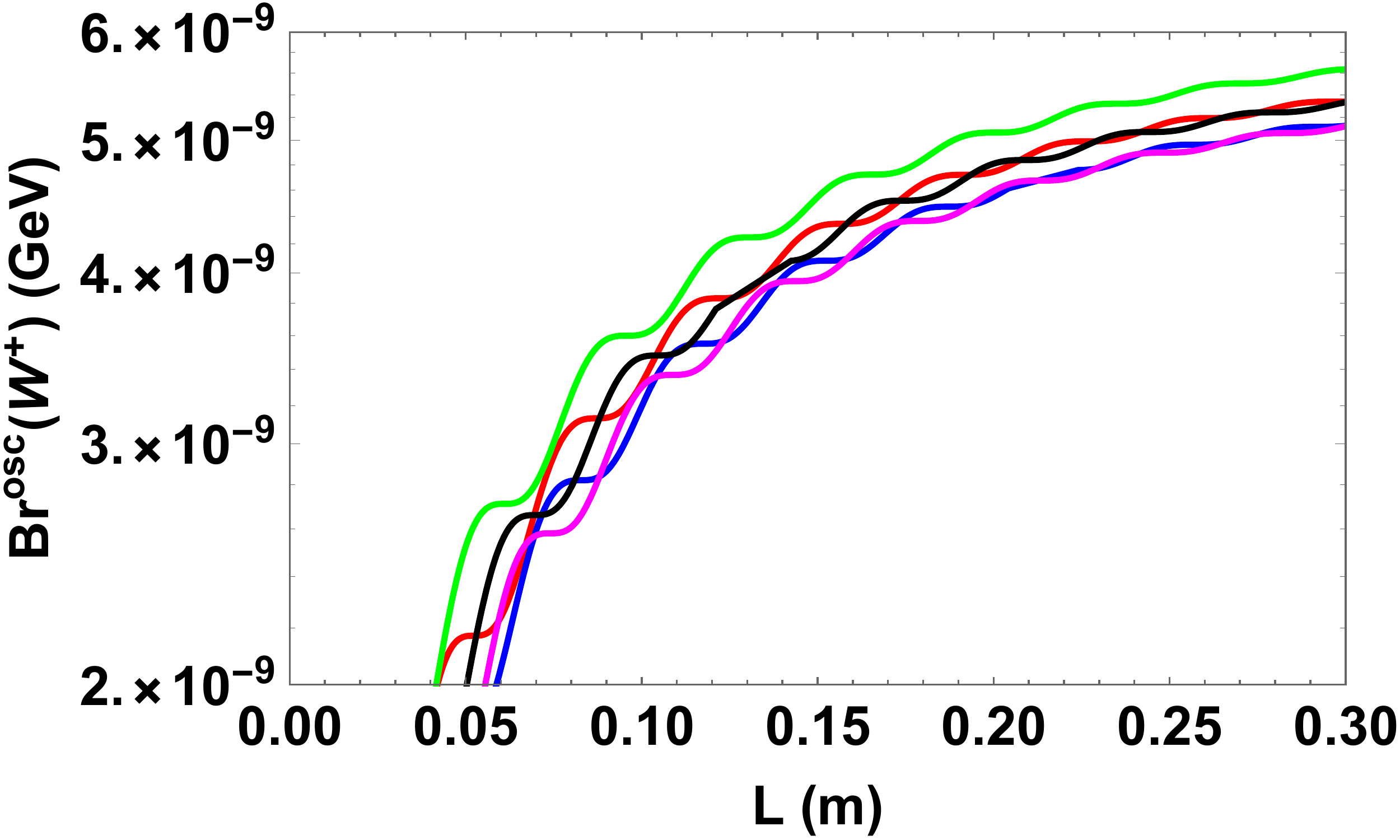} \\ \vspace{0.5 cm}
\includegraphics[scale = 0.28]{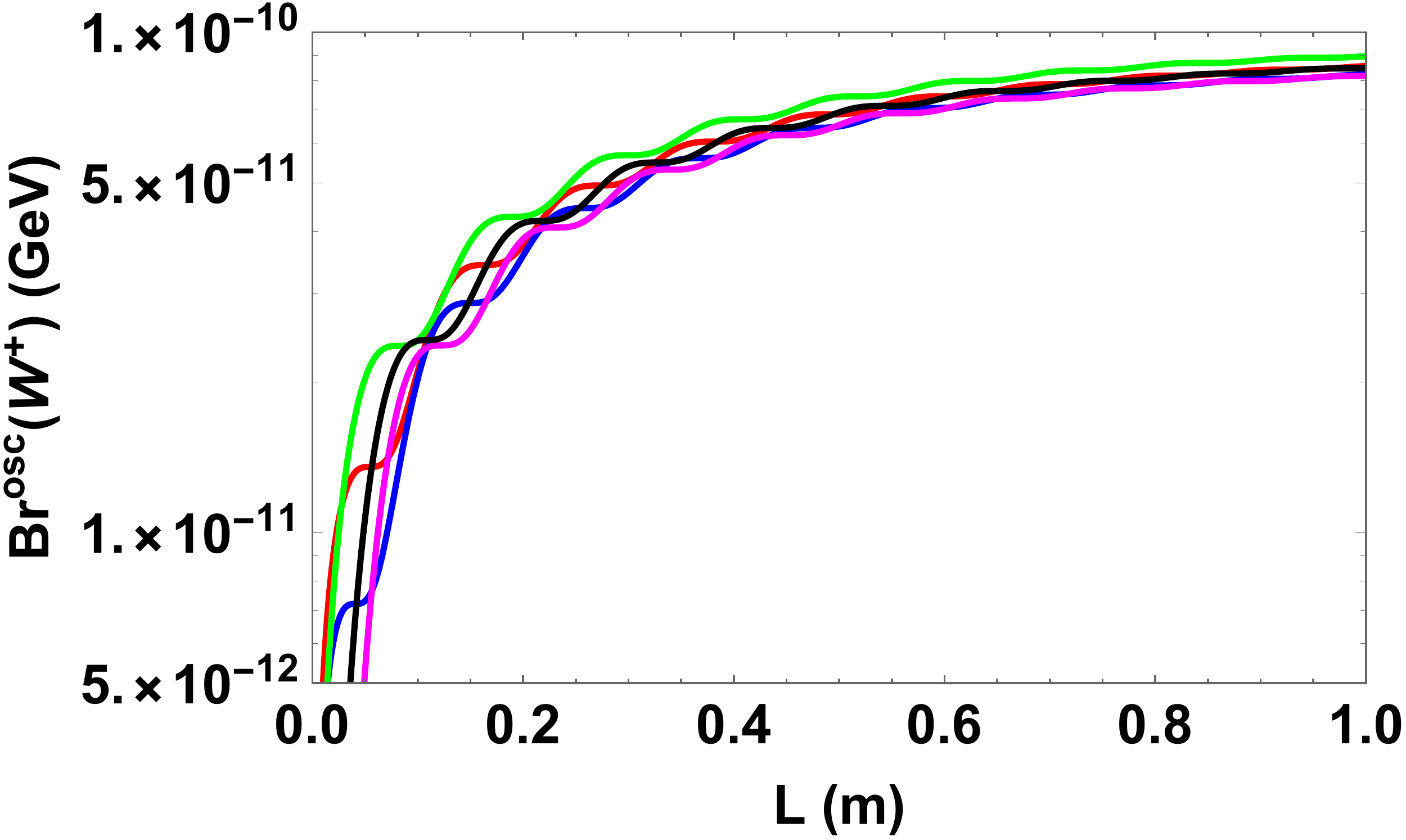}\hspace{0.3 cm}
\includegraphics[scale = 0.28]{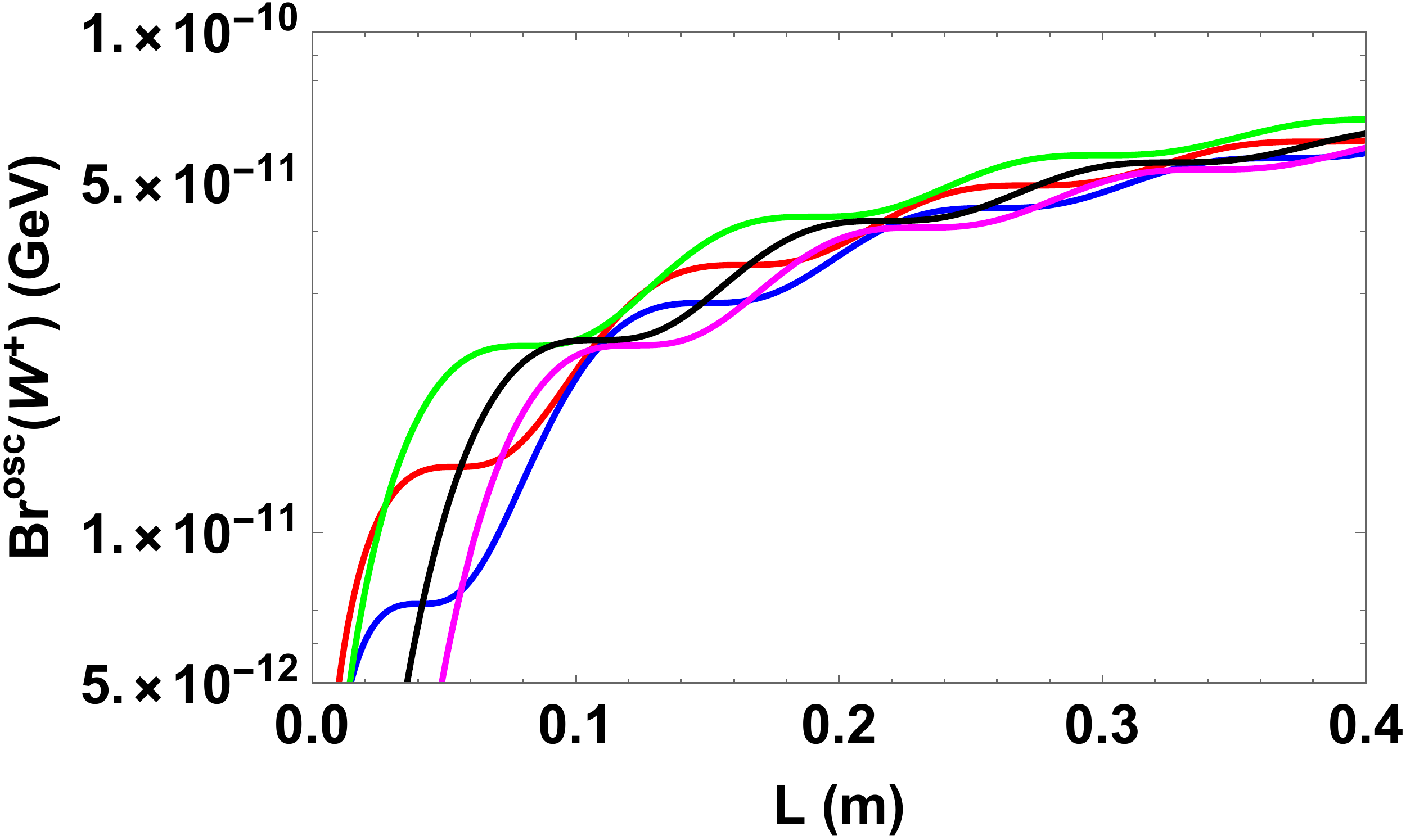}
\caption{Branching ratios ${\rm Br}^{{\rm osc}}(W^{+})$ for $\gamma_N \beta_N=2$ and $y=20$ as a function of $L$ for different $\theta$ angles. The red line stands for $\theta = 0$, blue for $\theta = \pi/4$, green for $\theta = \pi/2$, magenta for $\theta = 3 \pi /4$ and the black for $\theta = \pi$. Top: we provided $M_N=6$ GeV and $|B_{\ell N}|^2 \approx \mathcal{K}^{\rm Ma}/10=10^{-6}$. Bottom: we used $M_N=12$ GeV and $|B_{\ell N}|^2 \approx \mathcal{K}^{\rm Ma}/10=10^{-8}$.}
\label{fig:BroscWpL}
\end{figure}

\begin{figure}[H]
\centering
\includegraphics[scale = 0.42]{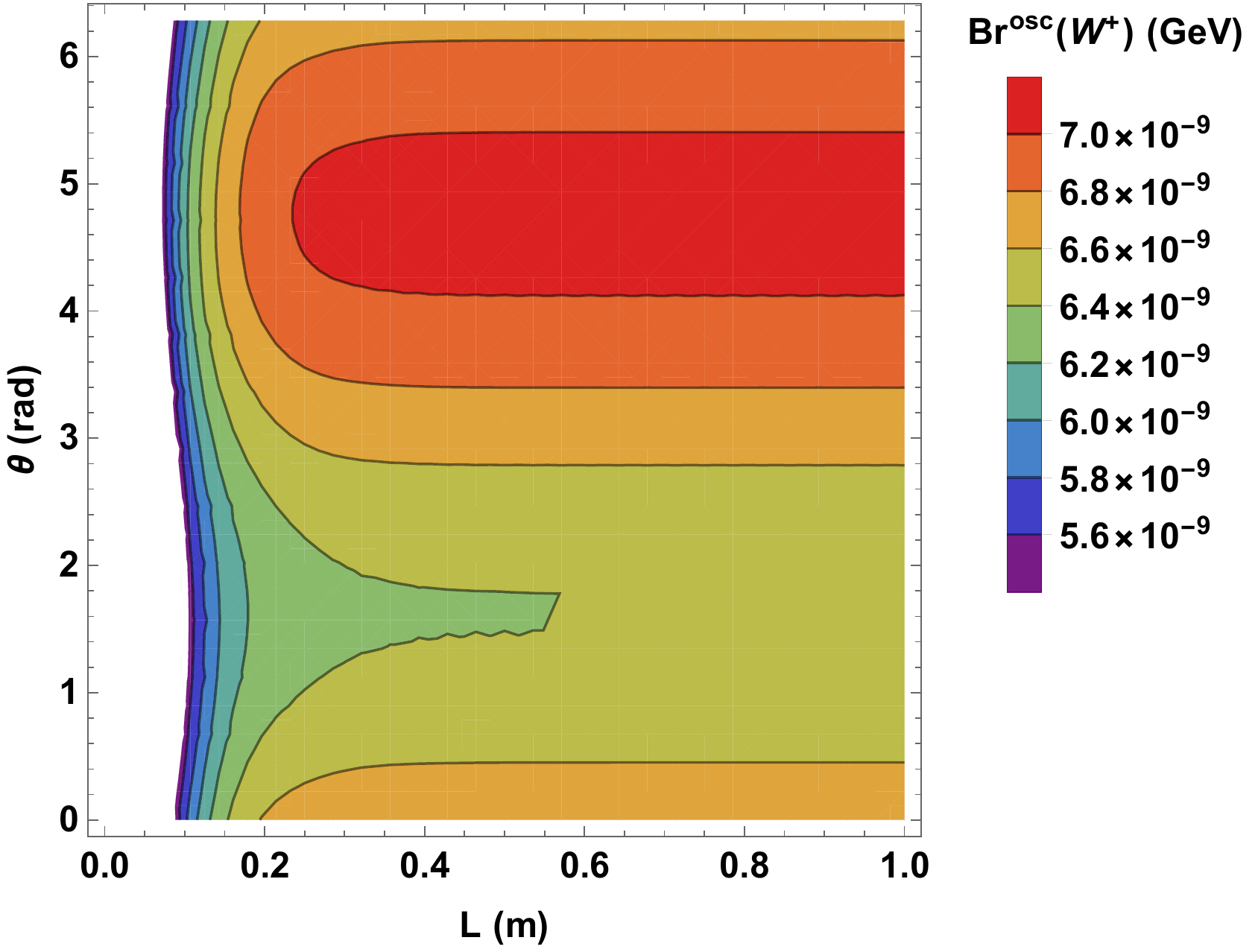}\hspace{0.3 cm}
\includegraphics[scale = 0.28]{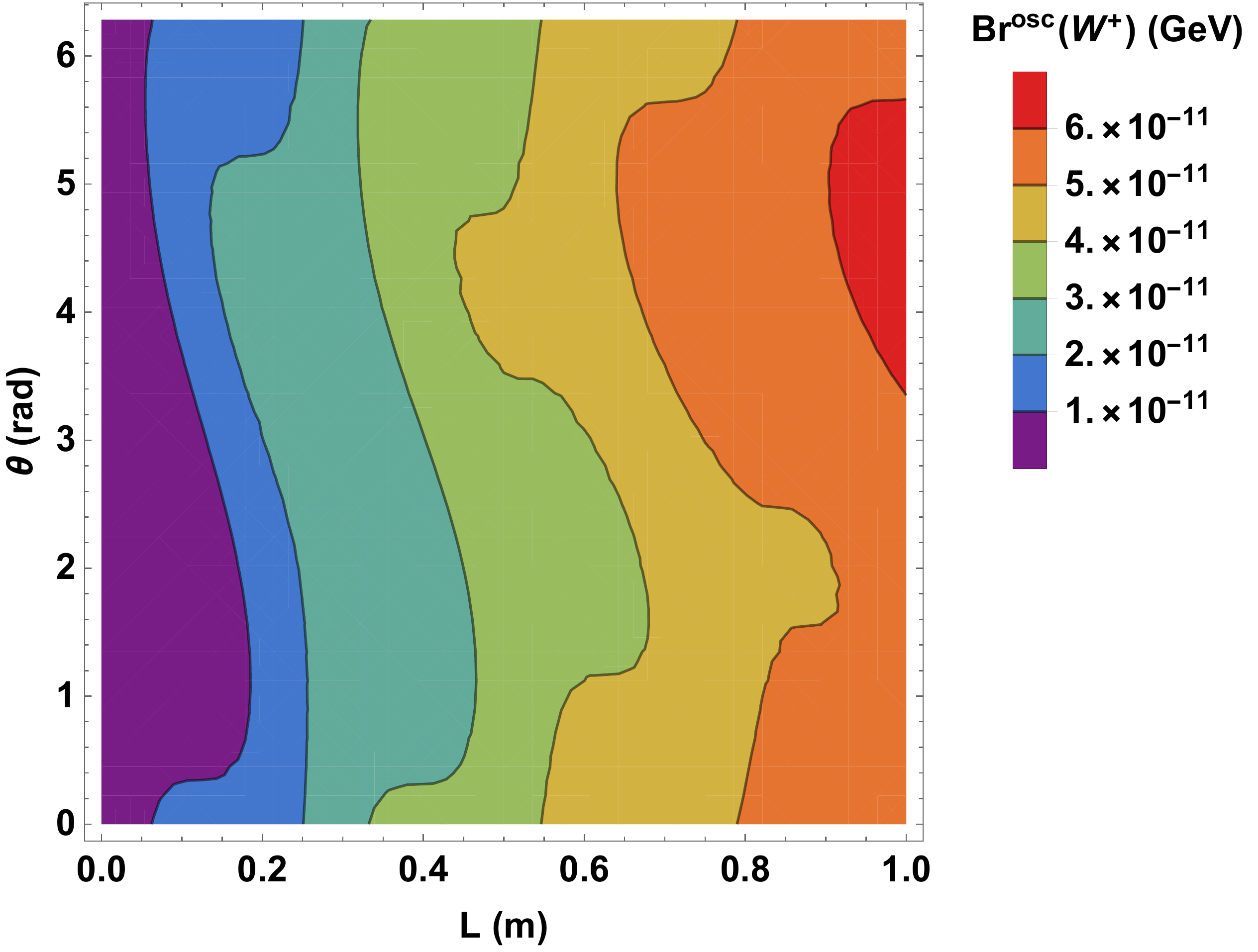}  \\ \vspace{0.5 cm}
\includegraphics[scale = 0.28]{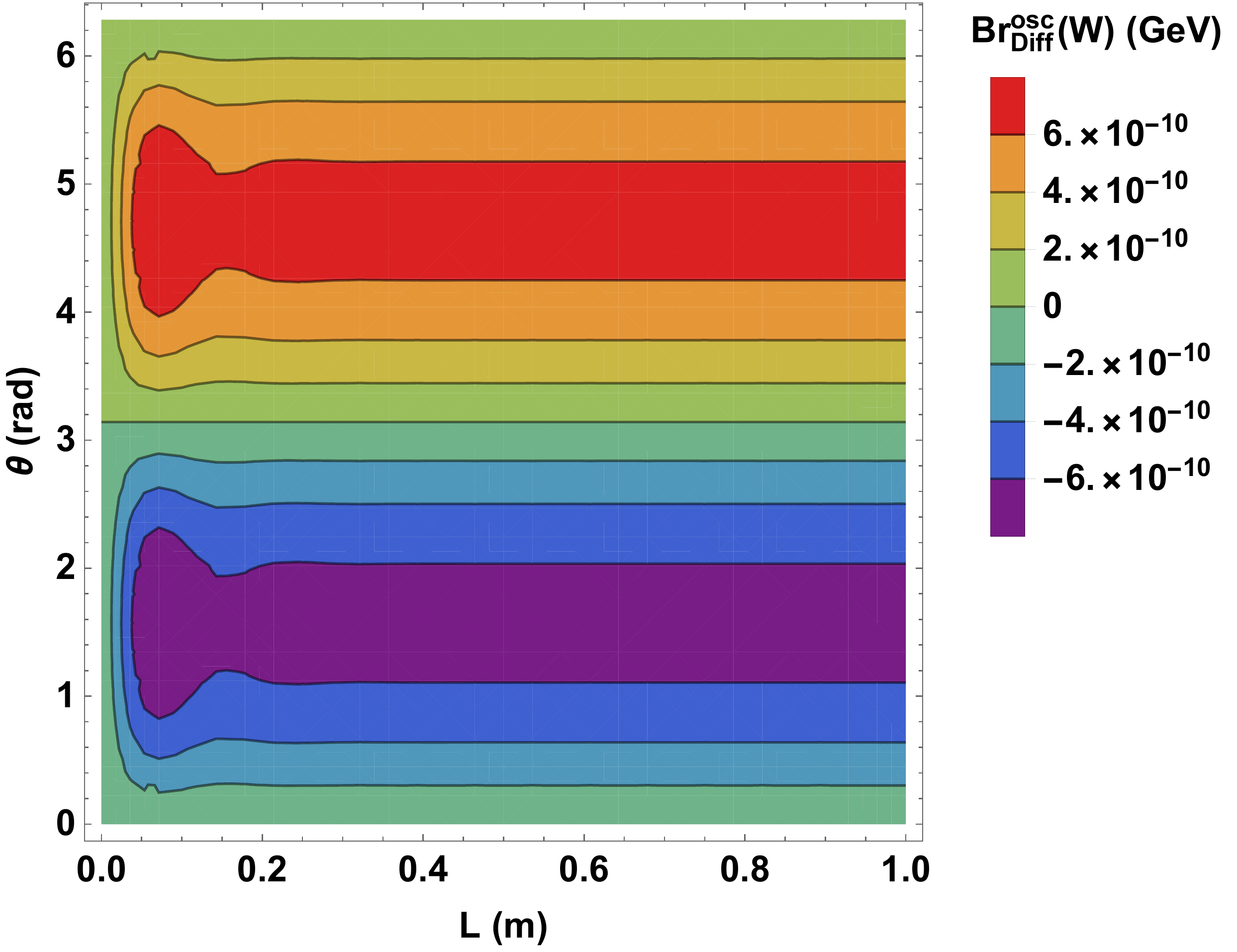} \hspace{0.3 cm}
\includegraphics[scale = 0.28]{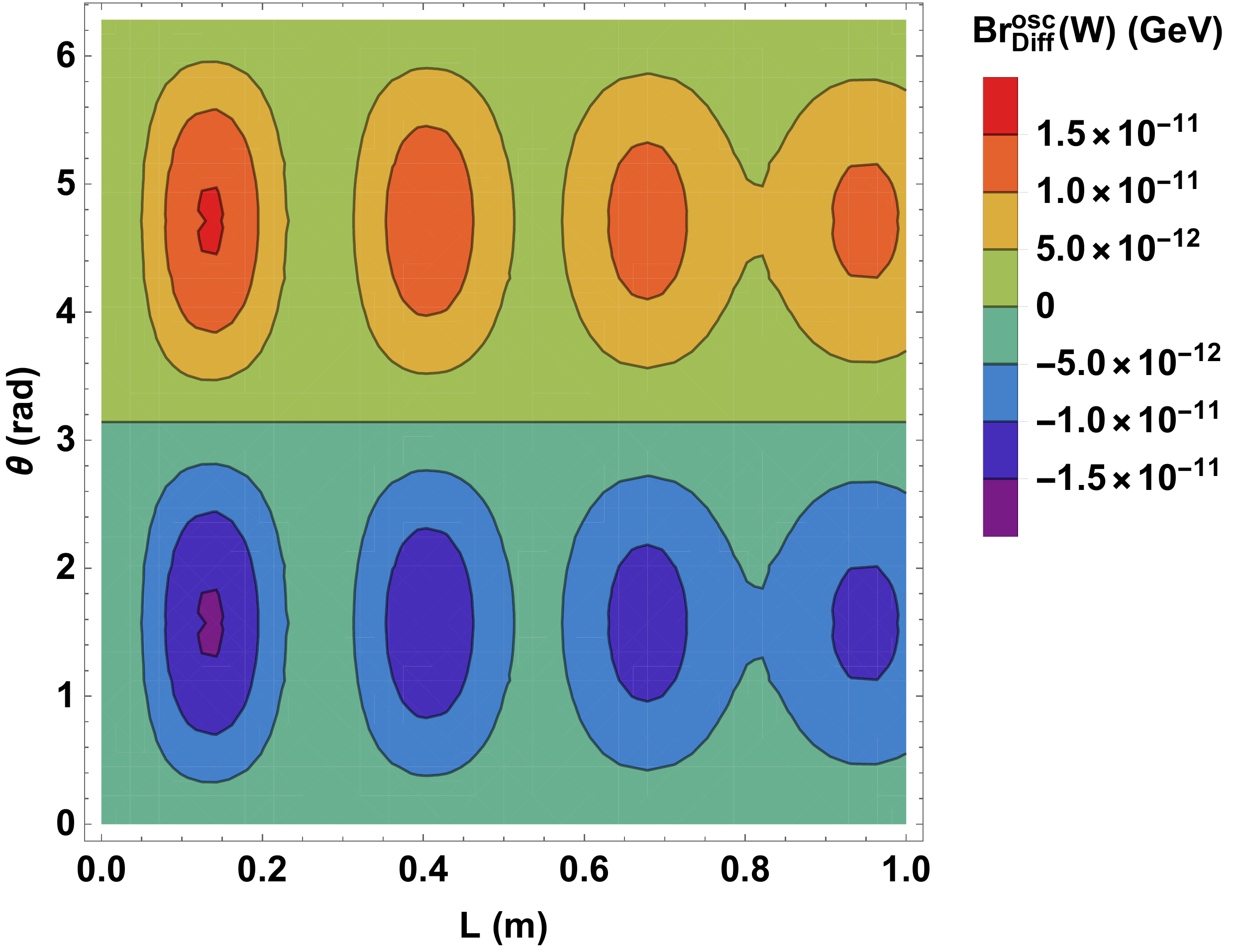}
\caption{Top: Branching ratios ${\rm Br}^{{\rm osc}}(W^{+})$ for $\gamma_N \beta_N=2$, $y=20$ as a function of $\theta$ and $L$; here we provided $M_N=7$ (GeV) and $|B_{\ell N}|^2 \approx \mathcal{K}^{\rm Ma}/10=10^{-6}$  (Left-panel) and $M_N=10$ (GeV) and $|B_{\ell N}|^2 \approx \mathcal{K}^{\rm Ma}/10=10^{-8}$ (Right-Panel). Bottom: Branching ratio difference ${\rm Br}_{\rm Diff}^{{\rm osc}}(W) \equiv {\rm Br}^{{\rm osc}}(W^{+})-{\rm Br}^{{\rm osc}}(W^{-})$ for $\gamma_N=2$, $\beta_N=1$, $y=20$ as a function of $\theta$ and $L$; here we provided $M_N=7$ (GeV) and $|B_{\ell N}|^2 \approx \mathcal{K}^{\rm Ma}/10=10^{-6}$ (Left-Panel) and $M_N=10$ (GeV) and $|B_{\ell N}|^2 \approx \mathcal{K}^{\rm Ma}/10=10^{-8}$ (Right-Panel).}
\label{fig:BroscWp2d}
\end{figure}

We note that in all these cases, the oscillation length $L_{\rm osc} = (2 \pi \gamma_N \beta_N/y) (1/\Gamma_N)$ is proportional to $1/\Gamma_N \propto 1/M_N^5$, i.e., it decreases fast when $M_N$ increases. We further note that at increased values of $M_N$ the oscillation effects, especially in the differential decay width, are more difficult to discern due to exponential attenuation. 
Fig.~\ref{fig:BroscWpL} shows the branching ratios as a function of the distance $L$ between the two vertices,  for $\gamma_N \beta_N=2$ and $y=20$, and for different fixed values of $\theta$. We notice that in these cases the modulation of  ${\rm Br}^{\rm osc}(W^+)$ could be possibly detected at HL-LHC for different angles when $L < 0.3$ m if $|B_{\ell N}|^2=10^{-6}$, and  $0.1 < L < 0.25$ m if $|B_{\ell N}|^2=10^{-8}$. 
In Fig.~\ref{fig:BroscWp2d} we show in the top two panels contour plots for branching ratios ${\rm Br}^{\rm osc}(W^+)$ as a function of $L$ and $\theta$ for various fixed values of $M_N$ and $|B_{\ell N}|^2$, and in the lower two panels the difference ${\rm Br}^{\rm osc}(W^+) - {\rm Br}^{\rm osc}(W^-)$.

We wish to point out that the considered number $\sim 10^{11}$ of produced $W^{\pm}$ in HL-LHC is probably an optimistic expectation; due to possible background and detector performance problems the effective number of $W^{\pm}$ may be significantly lower.

Finally, we point out that the possibility to detect heavy neutrino oscillations in rare top quark decays (cf.~Appendix~\ref{app4}) is low due to the relatively small number of expected produced top quarks to be collected in the future experiments.
\section{Acknowledgments}
This work was supported in part by FONDECYT Grant No.~3180032 (J.Z.S.) and FONDECYT Grant No.~1180344 (G.C.). The authors would like to thank Oliver Fischer for useful information regarding LHCb and Belle results.

\appendix
\section{ Kinematical function present in the amplitude}
\label{app1}
 The averaged squared amplitudes for general processes involving $n$ sterile neutrinos are
 {\nopagebreak
\begin{small}
 \begin{align}
\nonumber f_{\rm kin}^{\rm LV}&(k_1,p_1,p_2,p_{\ell^{'}},p_{\nu}) = \\
\nonumber &  -32 {m_{\ell^{'}}}^2 (p_{\ell^{'}} \cdot p_{\nu})^2 \Big[ 2 (k_1 \cdot p_1)(k_1 \cdot p_2)+M_W^2 (p_1 \cdot p_2) \Big] -16 {m_{\ell^{'}}}^2 (p_{\ell^{'}} \cdot p_{\nu})  \\
\nonumber & \times  \Big [ 2 (k_1 \cdot p_1) \Big( ({m_{\ell^{'}}}^2-2M_W^2)(k_1 \cdot p_2) - 2 \big ( (k_1 \cdot p_{\ell^{'}})+(k_1 \cdot p_{\nu}) \big) \big( (p_2 \cdot p_{\ell^{'}})+(p_2 \cdot p_{\nu}) \big) \Big) \Big ]  \\
\nonumber & + M_W^2 \Big[ ({m_{\ell^{'}}}^2-2 M_W^2)(p_1 \cdot p_2)-2\big( (p_1 \cdot p_{\ell^{'}})+(p_1 \cdot p_{\nu}) \big) \big( (p_2 \cdot p_{\ell^{'}})+(p_2 \cdot p_{\nu}) \big) \Big] \\
\nonumber &  + 32 M_W^2 \Big[M_W^2 \Big(-{m_{\ell^{'}}}^2 (p_1 \cdot p_{\nu}) 
\big( (p_2 \cdot p_{\ell^{'}})+2(p_2 \cdot p_{\nu}) \big) - ({m_{\ell^{'}}}^2-2 M_W^2)(p_1 \cdot p_{\ell^{'}}) (p_2 \cdot p_{\nu}) \Big) \\
& -2(k_1 \cdot p_1) \Big( ({m_{\ell^{'}}}^2-2M_W^2)(k_1 \cdot p_{\ell^{'}}) (p_2 \cdot p_{\nu}) + {m_{\ell^{'}}}^2 (k_1 \cdot p_{\nu}) \big( (p_2 \cdot p_{\ell^{'}})+2(p_2 \cdot p_{\nu}) \big) \Big) \Big] 
\label{kinLV}
\end{align}
\end{small}
\begin{small}
\begin{align}
\nonumber f^{\rm LC}_{\rm kin}&(k_1,p_1,p_2,p_{\ell^{'}},p_{\nu}) = \\
\nonumber& 16 \Bigg[ 2(k_1 \cdot p_1) \Bigg(  {m_{\ell^{'}}}^4 M_N^2 (k_1 \cdot p_2) (p_{\ell^{'}} \cdot p_{\nu})-2{m_{\ell^{'}}}^4 (k_1 \cdot p_N) (p_2 \cdot p_N) (p_{\ell^{'}} \cdot p_{\nu}) \\
\nonumber &- 2 {m_{\ell^{'}}}^2 M_{N}^2 M_W^2 (k_1 \cdot p_2) (p_{\ell^{'}} \cdot p_{\nu})+2 {m_{\ell^{'}}}^2 M_{N}^2 (k_1 \cdot p_{\nu})
 \Big[ M_W^2 \big[(p_2 \cdot p_{\ell^{'}})+2(p_2 \cdot p_{\nu}) \big] \\
 \nonumber &  -(p_{\ell^{'}} \cdot p_{\nu})\big[(p_2 \cdot p_{\ell^{'}})+(p_2 \cdot p_{\nu})\big] \Big]+2M_N^2 (k_1 \cdot p_{\ell^{'}}) \Big[ M_W^2 \big( {m_{\ell^{'}}}^2-2M_W^2 \big) (p_2 \cdot p_{\nu})\\
 \nonumber & -{m_{\ell^{'}}}^2 (p_{\ell^{'}} \cdot p_{\nu}) \big[(p_2 \cdot p_{\ell^{'}})+(p_2 \cdot p_{\nu}) \big] \Big]+2{m_{\ell^{'}}}^2 M_N^2 (k_1 \cdot p_2) (p_{\ell^{'}} \cdot p_{\nu})^2 \\
 \nonumber&-4{m_{\ell^{'}}}^2 M_W^2 (k_1 \cdot p_N) (p_2 \cdot p_{\nu}) (p_{\ell^{'}} \cdot p_N)+4{m_{\ell^{'}}}^2 M_W^2 (k_1 \cdot p_N) (p_2 \cdot p_N) (p_{\ell^{'}} \cdot p_{\nu})\\
 \nonumber & +4{m_{\ell^{'}}}^2 (k_1 \cdot p_N) (p_N \cdot p_{\nu}) \Big[ (p_{\ell^{'}} \cdot p_{\nu}) \big( (p_{2} \cdot p_{\ell^{'}})+(p_{2} \cdot p_{\nu}) \big)-M_W^2 \big((p_{2} \cdot p_{\ell^{'}})+2(p_{2} \cdot p_{\nu}) \big) \Big] \\
 \nonumber&-4{m_{\ell^{'}}}^2 (k_1 \cdot p_N) (p_2 \cdot p_N) (p_{\ell^{'}} \cdot p_{\nu})^2+4{m_{\ell^{'}}}^2 (k_1 \cdot p_N) (p_2 \cdot p_{\ell^{'}}) (p_{\ell^{'}} \cdot p_N) (p_{\ell^{'}} \cdot p_{\nu}) \\
 \nonumber &+4{m_{\ell^{'}}}^2 (k_1 \cdot p_N) (p_2 \cdot p_{\nu}) (p_{\ell^{'}} \cdot p_N) (p_{\ell^{'}} \cdot p_{\nu})+8M_W^4 (k_1 \cdot p_N) (p_2 \cdot p_{\nu}) (p_{\ell^{'}} \cdot p_N)  \Bigg) \\
 \nonumber&+ M_W^2 \Bigg( {m_{\ell^{'}}}^4 M_N^2 (p_1 \cdot p_2) (p_{\ell^{'}} \cdot p_{\nu}) -2 {m_{\ell^{'}}}^4 (p_1 \cdot p_N) (p_2 \cdot p_N) (p_{\ell^{'}} \cdot p_{\nu}) \\
 \nonumber &-2 {m_{\ell^{'}}}^2 M_N^2 M_W^2 (p_1 \cdot p_2) (p_{\ell^{'}} \cdot p_{\nu}) +2 {m_{\ell^{'}}}^2 M_N^2 (p_1 \cdot p_{\nu})
 \Big[ M_W^2 \big( (p_2 \cdot p_{\ell^{'}})+2(p_2 \cdot p_{\nu})  \big) \\
 \nonumber & -(p_{\ell^{'}} \cdot p_{\nu})\big( (p_2 \cdot p_{\ell^{'}})+(p_2 \cdot p_{\nu}) \big)  \Big] + 2M_N^2(p_1 \cdot p_{\ell^{'}}) \Big[M_W^2\big( {m_{\ell^{'}}}^2-2M_W^2 \big)(p_2 \cdot p_{\nu}) \\
 \nonumber& - {m_{\ell^{'}}}^2(p_{\ell^{'}} \cdot p_{\nu}) \big( (p_2 \cdot p_{\ell^{'}})+(p_2 \cdot p_{\nu}) \big) \Big] + 2 {m_{\ell^{'}}}^2 M_N^2 (p_1 \cdot p_2) (p_{\ell^{'}} \cdot p_{\nu})^2 \\
 \nonumber &-4{m_{\ell^{'}}}^2 M_W^2 (p_1 \cdot p_N) (p_2 \cdot p_{\nu}) (p_{\ell^{'}} \cdot p_N) + 4{m_{\ell^{'}}}^2 M_W^2 (p_1 \cdot p_N)
 (p_2 \cdot p_N) (p_{\ell^{'}} \cdot p_{\nu})\\
 \nonumber &+ 4 {m_{\ell^{'}}}^2 (p_1 \cdot p_N) (p_N \cdot p_{\nu}) \Big[ (p_{\ell^{'}} \cdot p_{\nu}) \big( (p_2 \cdot p_{\ell^{'}})+(p_2 \cdot p_{\nu}) \big)-M_W^2 \big( (p_2 \cdot p_{\ell^{'}})+2 (p_2 \cdot p_{\nu}) \big) \Big] \\
 \nonumber & -4 {m_{\ell^{'}}}^2 (p_1 \cdot p_N) (p_2 \cdot p_N) (p_{\ell^{'}} \cdot p_{\nu})^2+4{m_{\ell^{'}}}^2 (p_1 \cdot p_N) (p_2 \cdot p_{\ell^{'}}) (p_{\ell^{'}} \cdot p_N) (p_{\ell^{'}} \cdot p_{\nu}) \\
  & +4 {m_{\ell^{'}}}^2 (p_1 \cdot p_N) (p_2 \cdot p_{\nu}) (p_{\ell^{'}} \cdot p_N) (p_{\ell^{'}} \cdot p_{\nu})+8 M_W^4 (p_1 \cdot p_N) (p_2 \cdot p_{\nu}) (p_{\ell^{'}} \cdot p_N) \Bigg) \Bigg]  \ .
\label{kinLC}
\end{align}
\end{small}
}
\section{ Delta Function Approximation for the Propagator Product}
\label{app2}
\begin{figure}[H]
\centering
\includegraphics[scale = 0.65]{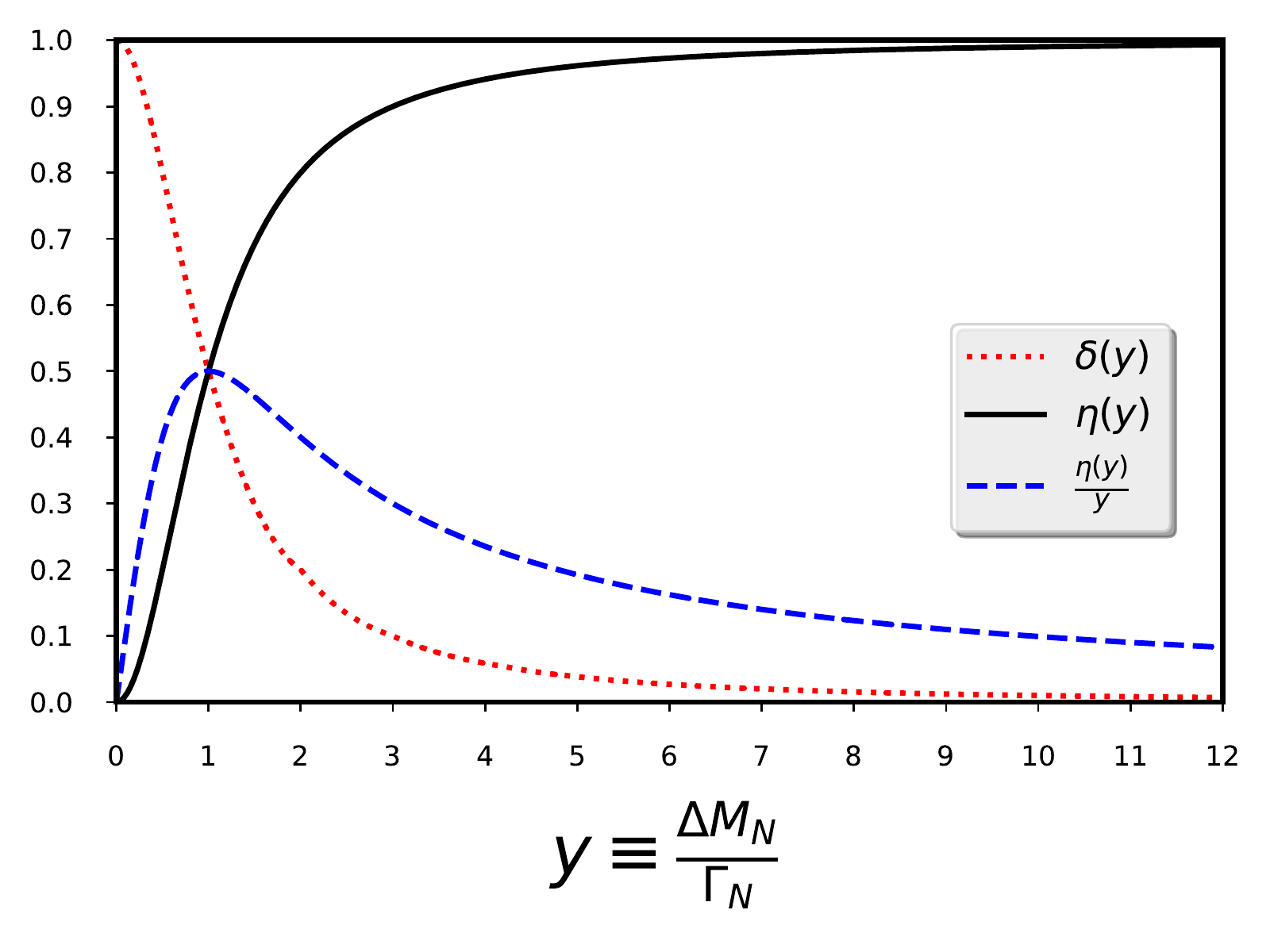}
\caption{Solid line (online red) overlap function $\delta_{12}$. Dashed line (online blue) $\eta(y)$ function. Dotted line (online black) $\eta(y)/y$ function. Figure taken from \cite{Zamora-Saa:2016ito}.}
\label{delta_eta}
\end{figure}

The off-diagonal elements in Eq.~(\ref{dw}) can be separated in two different terms, one containing $\cos (\theta_X)$ and the other one $\sin (\theta_X)$. The former is proportional to the factor $\delta_{12}$ which was discussed in Sec.~\ref{s3} after Eq.~\eqref{DWE} and the latter is proportional to the factor $\eta(y)/y$ appearing in the expression $\Omega$ which is the imaginary part of a combination of the propagators of $N_1$ and $N_2$
\begin{align}
\label{A1}
\Omega &= \frac{
\left( p_N^2 - M_{N_1}^2 \right)  \Gamma_{N_2} M_{N_2}
- \Gamma_{N_1} M_{N_1} \left( p_N^2 - M_{N_2}^2 \right)
}
{\left[ \left( p_N^2 - M_{N_1}^2 \right)^2 + \Gamma_{N_1}^2 M_{N_1}^2
\right]
\left[ \left( p_N^2 - M_{N_2}^2 \right)^2 + \Gamma_{N_2}^2 M_{N_2}^2
\right]}\\
\label{A2}
& \approx  \eta(y) \; \frac{\pi}{M^{2}_{N_2}-M^{2}_{N_1}} \left [
\delta  ( p^{2}_{N}-M^{2}_{N_2})+ \delta  ( p^{2}_{N}-M^{2}_{N_1})  \right ] \ .
\end{align}
In the narrow width approximation (NWA), $\eta(y)=1$. NWA is valid only when  $\Gamma_N \equiv (\Gamma_{N_1} + \Gamma_{N_2})/2 \ll |\Delta M_N|$ (where $\Delta M_N \equiv M_{N_2} - M_{N_1}$). It turns out that
\begin{equation}
\eta(y) = \frac{y^2}{1 + y^2} ,
\end{equation}
where $y \equiv |\Delta M_N|/\Gamma_N$. The values of $\eta(y)$ were evaluated numerically in previous works \cite{Cvetic:2013eza,Cvetic:2014nla,Zamora-Saa:2016qlk,Zamora-Saa:2016ito} using finite $\Delta M_N$, and in Ref.~\cite{Cvetic:2015naa} the analitical expresion. The values of $\eta(y)$ are presented in Fig.~\ref{delta_eta} as a function of $y \equiv |\Delta M_N|/\Gamma_N$. 
\section{ Kinematical function for numerical integration}
\label{app3}
Numerical integration functions mentioned in Sec.~\ref{s3} are
\bea
z_{\rm max}(E_{\ell^{'}}) = z(|{\vec p}_{\nu}|_{\rm min}; |{\vec p}_{\ell^{'}}|) 
& = &  +\frac{1}{\Gamma_W M_W}( M_W^2 + M_N^2 - 2 M_N E_{\ell^{'}}- {m_2}^2) 
\nonumber\\ &&
 - \frac{M_N}{\Gamma_W M_W} \left[ 1- \frac{{m_2}^2}{(M_N^2 - 2 M_N E_{\ell^{'}} + {m_{\ell^{'}}}^2)} \right]\nonumber\\
 && (M_N - E_{\ell^{'}} - \sqrt{ E_{\ell^{'}}^2 - {m_{\ell^{'}}}^2}),
\label{zmax}
\\
z_{\rm min}(E_{\ell^{'}}) = z(|{\vec p}_{\nu}|_{\rm max}; |{\vec p}_{\ell^{'}}|) 
& = &  +\frac{1}{\Gamma_W M_W}( M_W^2 + M_N^2 - 2 M_N E_{\ell^{'}}- {m_2}^2 )  
\nonumber\\ &&
- \frac{M_N}{\Gamma_W M_W} \left[ 1- \frac{{m_2}^2}{(M_N^2 - 2 M_N E_{\ell^{'}} + {m_{\ell^{'}}}^2)} \right] \nonumber\\
&&(M_N - E_{\ell^{'}} + \sqrt{ E_{\ell^{'}}^2 - {m_{\ell^{'}}}^2}),
\label{zmin}
\eea
and the (dimensionless) coefficients ${\bar A}_j$ ($j=1,2$) in the integrand of Eq.~(\ref{barInt}) are
\bea
{\bar A}_2 & = & -\frac{4 \Gamma_W {m_{\ell^{'}}}^2}{M_N^2 M_W^3} ( M_N^2 + {m_{\ell^{'}}}^2),
\label{barA2}
\\
\nonumber {\bar A}_1 & = &  \frac{4 {m_{\ell^{'}}}^2}{M_N^2 M_W^4} \left[ M_N^2 ( 2 M_W^2 - M_N^2 - {m_{\ell^{'}}}^2) + {m_2}^2 (- 2 M_W^2 + 2 M_N^2 - {m_2}^2 - {m_{\ell^{'}}}^2) \right], \\
\label{barA1}
\eea
and the (dimensionless) function ${\bar A}_0(E_{\ell^{'}})$ in the integrand of Eq.~(\ref{barInt}) is 
\bea
{\bar A}_0(E_{\ell^{'}}) & = & - \frac{64}{\Gamma_W M_W} E_{\ell^{'}}^2 + \frac{32}{\Gamma_W M_W^3 M_N} \big[ M_W^2 M_N^2 \nonumber \\
&& - {m_2}^2(M_W^2 + {m_{\ell^{'}}}^2) + {m_{\ell^{'}}}^2 (M_W^2 + M_N^2) \big] E_{\ell^{'}}+ {\bar A}_0(0),
\label{barA0}
\eea
where the constant ${\bar A}_0(0)$ is
\bea
{\bar A}_0(0) & = & \frac{4 {m_{\ell^{'}}}^2}{\Gamma_W M_W^5 M_N^2} 
{\big [}  - M_N^2 M_W^2  (M_W^2 + 3 M_N^2 - \Gamma_W^2)  - {m_{\ell^{'}}}^2 M_N^2 (3 M_W^2+M_N^2)  
\nonumber\\ &&
+ {m_2}^2 M_W^2 (3 M_W^2 + 6 M_N^2 +\Gamma_W^2) + {m_2}^2 {m_{\ell^{'}}}^2 (M_W^2 + 2 M_N^2) - \nonumber \\
&&{m_2}^4 (3 M_W^2 +{m_{\ell^{'}}}^2) {\big ]}. 
\label{barA00}
\eea
\section{ Top quark Decays}
\label{app4}
\begin{figure}[H]
\centering
\includegraphics[scale = 0.66]{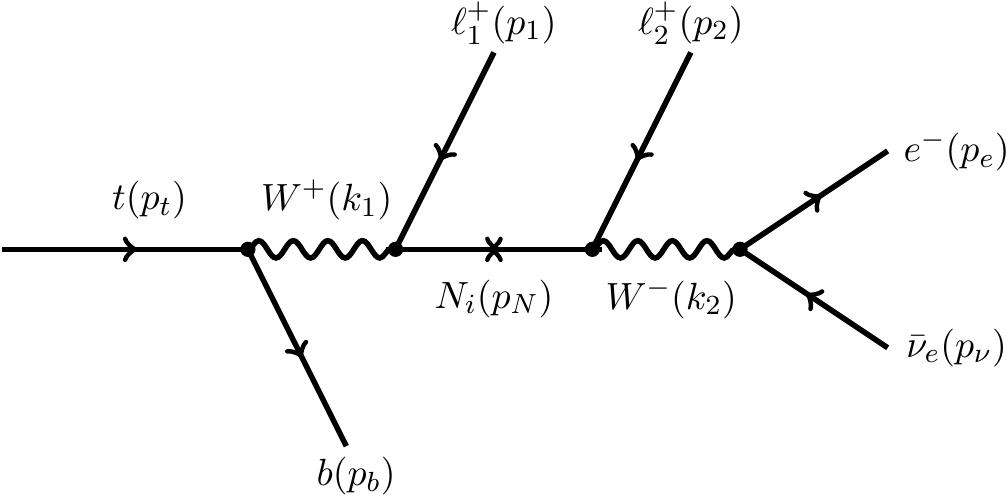}
\includegraphics[scale = 0.66]{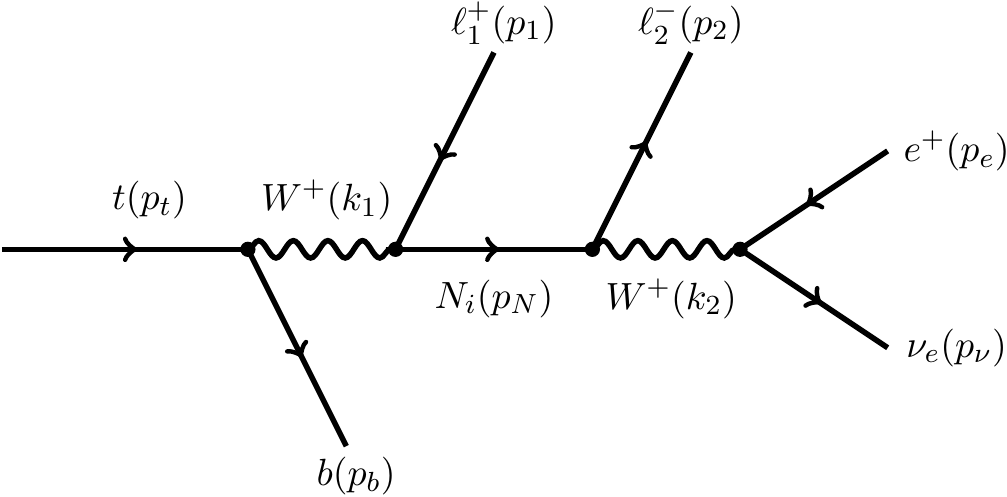}
\caption{Feynman diagrams for the LNV process $t \rightarrow \ell_1^+ \ell_2^+ e^- \bar{\nu}_{e}$ (left panel) and LNC process $t \rightarrow \ell_1^+ \ell_2^- e^+ \nu_{e}$ (right panel).}
\label{fig:tdecays}
\end{figure}

The relevant Feynman diagrams the rare top-quark decay are presented in Fig.~\ref{fig:tdecays}
By means of Eq.~\eqref{dw} and Eq.~\eqref{DWE},  the top quark decay width can be expressed as
\begin{subequations}
\begin{align}
\nonumber \Gamma\big( t(p_t) \to & b(p_b) \ell^+_1(p_1) \ell^+_2(p_2) e^-(p_e) \bar{\nu}_1(p_{\nu}) \big) \equiv \Gamma_{LV}(t^{+}) = \\
&= \Gamma\big( t(p_t) \to  b(p_b) W^+({k_1}) \big) \; \frac{\Gamma_{LV}\big( W^+({k_1}) \to \ell^+_1(p_1) \ell^+_2(p_2) e^-(p_e) \bar{\nu}_e(p_{\nu}) \big)}{\Gamma\big( W^+(k_1) \to {\rm all} \big)} \, \\
\nonumber \Gamma\big( t(p_t) \to & b(p_b) \ell^+_1(p_1) \ell^-_2(p_2) e^+(p_e) \nu_1(p_{\nu}) \big) \equiv \Gamma_{LC}(t^{+}) = \\
&= \Gamma\big( t(p_t) \to  b(p_b) W^+({k_1}) \big) \; \frac{\Gamma_{LC}\big( W^+({k_1}) \to \ell^+_1(p_1) \ell^-_2(p_2) e^+(p_e) \nu_e(p_{\nu}) \big)}{\Gamma\big( W^+(k_1) \to {\rm all} \big)} \ .
\end{align}
\end{subequations}
For the charge-conjugate (top antiquark) decays we have
\begin{subequations}
\begin{align}
\nonumber \Gamma\big( \bar{t}(p_t) \to & \bar{b}(p_b) \ell^-_1(p_1) \ell^-_2(p_2) e^+(p_e) {\nu}_e(p_{\nu}) \big) \equiv \Gamma_{LV}(t^{-}) = \\
&= \Gamma\big( \bar{t}(p_t) \to  \bar{b}(p_b) W^-({k_1}) \big) \; \frac{\Gamma_{LV}\big( W^-({k_1}) \to \ell^-_1(p_1) \ell^-_2(p_2) e^+(p_e){\nu}_e(p_{\nu}) \big)}{\Gamma\big( W^-(k_1) \to {\rm all} \big)} \, \\
\nonumber \Gamma\big( \bar{t}(p_t) \to & \bar{b}(p_b) \ell^-_1(p_1) \ell^+_2(p_2) e^-(p_e) \bar{\nu}_e(p_{\nu}) \big) \equiv \Gamma_{LC}(t^{-}) = \\
&= \Gamma\big( \bar{t}(p_t) \to  \bar{b}(p_b) W^-({k_1}) \big) \; \frac{\Gamma_{LC}\big( W^-({k_1}) \to \ell^-_1(p_1) \ell^+_2(p_2) e^-(p_e) \bar{\nu}_e(p_{\nu}) \big)}{\Gamma\big( W^-(k_1) \to {\rm all} \big)} \ ,
\end{align}
\end{subequations}
where the top decay width is given by \cite{Jezabek:1988iv}
\begin{align}
\nonumber \Gamma\big( t(p_t) \to  & b(p_b) W^+({k_1}) \big) = \Gamma\big(\bar{t}(p_t) \to  \bar{b}(p_b) W^-({k_1}) \big) \\
=& \frac{G_F m_t^3 |V_{tb}|^2}{8 \pi \sqrt{2}} \; \Bigg( 1-\frac{M_W^2}{m_t^2} \Bigg)^2 \Bigg( 1+\frac{2 M_W^2}{m_t^2} \Bigg) \Bigg[ 1-\frac{2 \alpha_s}{3 \pi}\Bigg( \frac{2 \pi^2}{3} -\frac{5}{2} \Bigg) \Bigg] \ .
\end{align}
By means the same procedure followed in sec.~\ref{s3} the top and anti-top decay width can be written as
\begin{small}
\begin{align}
\label{osctfulldw}
\nonumber \Gamma_{X}^{\rm osc}(t^{\pm})& = F\big[t(p_t)\big]  \Bigg[ \Bigg(1-\exp\Big[-\frac{L \ \Gamma_{\rm Ma}(M_N)}{\gamma_N \ \beta_N}\Big] \Bigg) \sum_{i=1}^2 \frac{|B_{\mu N_i}|^2 |B_{\tau N_i}|^2}{\Gamma_{\rm Ma}(M_{N})}  \\
\nonumber & +\frac{2 |B_{\mu N_1}| |B_{\tau N_1}| |B_{\mu N_2}| |B_{\tau N_2}|  L_{\rm osc}}{\Gamma_{\rm Ma}(M_N)^2 L_{\rm osc}^2+4 \beta_N^2 \gamma_N^2 \pi^2} \times \\
\nonumber &  \quad  \Bigg(  \Gamma_{\rm Ma}(M_N) L_{\rm osc} \cos(\theta_X) \mp 2 \beta_N \gamma_N \pi \sin(\theta_X) + \exp\Big[-\frac{L \ \Gamma_{\rm Ma}(M_N)}{\gamma_N \ \beta_N}\Big] \times \\
& \quad  \Bigg[2 \beta_N \gamma_N \pi \sin\Big(2\pi \; \frac{L}{L_{\rm osc}} \pm \theta_{X} \Big)  
-\Gamma_{\rm Ma}(M_N) L_{\rm osc}\cos \Big(2\pi \; \frac{L}{L_{\rm osc}} \pm \theta_{X} \Big) \Bigg]  \Bigg) 
\Bigg] \ .
\end{align}
\end{small}
Here, ($X=LV,\ LC$) and the function $F\big[t(p_t)\big]$ is given by
\begin{align}
F\big[t(p_t)\big] &= \Gamma\big( t(p_t) \to  b(p_b) W^+({k_1}) \big)\;  \frac{\widetilde{\Gamma}\big( W^+ \to \ell^+_1 N \big) \ \widetilde{\Gamma}\big( N \to \ell^+_2 e^- \bar{\nu}  \big)}{\Gamma\big( W^+(k_1) \to {\rm all} )}  \ ,
\end{align}
Therefore, the oscillating branching ratios are
\begin{align}
\label{osctopfullBR}
{\rm Br}_{X}^{{\rm osc}}(t^{\pm})& = \frac{\Gamma_{X}^{\rm osc}(t^{\pm})}{\Gamma(t^{+} \to {\rm all})} \ .
\end{align}


\end{document}